\documentclass{aa}

\usepackage{allrunes}
\usepackage{graphicx}
\usepackage{txfonts}
\usepackage{wasysym}
\usepackage{array}
\usepackage[pdftitle={Transit least-squares survey -- II. -- Discovery and validation of 17 new sub- to super-Earth-sized planets in multi-planet systems from {\it K2}}, colorlinks = true, breaklinks = true, citecolor = blue, linkcolor = blue, urlcolor = blue, pdfauthor = {Ren\'{e} Heller}]{hyperref}
\usepackage{esvect}  

\newcolumntype{C}[1]{>{\centering\let\newline\\\arraybackslash\hspace{0pt}}m{#1}}

\defcitealias{2019A&A...625A..31H}{Paper~I}
\newcommand{\PaperI}{\citetalias{2019A&A...625A..31H}}

\begin{document}

\title{Transit least-squares survey}
\subtitle{II. Discovery and validation of 17 new sub- to super-Earth-sized planets\\in multi-planet systems from {\it K2}}
\titlerunning{Transit Least-Squares Survey -- II. Discovery and validation of 17 new planets in {\it K2}}
\author{Ren\'{e} Heller\inst{1}
\and
Michael Hippke\inst{2}
\and
Kai Rodenbeck\inst{3,1}
          }

   \institute{Max Planck Institute for Solar System Research, Justus-von-Liebig-Weg 3, 37077 G\"ottingen, Germany, \href{mailto:heller@mps.mpg.de}{heller@mps.mpg.de}
   \and
   Sonneberg Observatory, Sternwartestra{\ss}e 32, 96515 Sonneberg, Germany, \href{mailto:michael@hippke.org}{michael@hippke.org}
   \and
   \scalebox{.983}[1.0]{Inst. for Astrophysics, Georg-August-Univ. G\"ottingen, Friedrich-Hund-Platz 1, 37077 G\"ottingen, Germany, \href{mailto:rodenbeck@mps.mpg.de}{rodenbeck@mps.mpg.de}}
             }

   \date{Received 2 April 2019; Accepted 13 May 2019}


\abstract{The extended {\it Kepler} mission ({\it K2}) has revealed more than 500 transiting planets in roughly 500\,000 stellar light curves. All of these were found either with the {\tt box least-squares} algorithm or by visual inspection.
Here we use our new {\tt transit least-squares} ({\tt TLS}) algorithm to search for additional planets around all {\it K2} stars that are currently known to host at least one planet.
We discover and statistically validate 17 new planets with radii ranging from about 0.7 Earth radii ($R_\oplus$) to roughly $2.2\,R_\oplus$ and a median radius of $1.18\,R_\oplus$. EPIC\,201497682.03, with a radius of $0.692_{-0.048}^{+0.059}\,R_\oplus$, is the second smallest planet ever discovered with {\it K2}. The transit signatures of these 17 planets are typically 200\,ppm deep (ranging from 100\,ppm to 2000\,ppm), and their orbital periods extend from about 0.7\,d to 34\,d with a median value of about 4\,d. Fourteen of these 17 systems only had one known planet before, and they now join the growing number of multi-planet systems. Most stars in our sample have subsolar masses and radii. The small planetary radii in our sample are a direct result of the higher signal detection efficiency that {\tt TLS} has compared to box-fitting algorithms in the shallow-transit regime. Our findings help in populating the period-radius diagram with small planets. Our discovery rate of about 3.7\,\% within the group of previously known {\it K2} systems suggests that {\tt TLS} can find over 100 additional Earth-sized planets in the data of the {\it Kepler} primary mission.}



   \maketitle
%

\section{Introduction}
\label{sec:introduction}

After the continuous monitoring of its prime observing field from 2009 to 2013, the repurposed {\it Kepler} telescope \citep{2010Sci...327..977B} performed another 19 campaigns of different target fields along the ecliptic, covered for about 75\,d each from 2014 until 2018 \citep{2014PASP..126..398H}. Although the failure of the second of the four reaction wheels of {\it Kepler} initially produced degraded photometric precision of the {\it K2} mission, new decorrelation techniques between photometrically extracted light curves and the telescope pointing \citep{2014PASP..126..948V,2015ApJ...806...30L,2016MNRAS.459.2408A} led to substantial improvements of the noise properties of the light curves. Finally, \citet{2016AJ....152..100L} constructed {\tt EVEREST}, an automated {\it K2} photometric extraction pipeline based on a pixel level decorrelation technique \citep{2015ApJ...805..132D} in combination with a Gaussian process optimization \citep{2015MNRAS.447.2880A}.

These improvements of the {\it K2} stellar photometry calibration allowed the confirmation of 359 exoplanets and 472 more candidates\footnote{\href{https://exoplanetarchive.ipac.caltech.edu/docs/counts_detail.html}{https://exoplanetarchive.ipac.caltech.edu/docs/counts\_detail.html} on 22 February 2019}, in addition to the 2331 confirmed planets and 2425 candidates yet to be confirmed from the {\it Kepler} primary mission. All {\it K2} planets and candidates were found using either a direct application of the {\tt box least-squares} ({\tt BLS}) \citep{2002A&A...391..369K} transit search algorithm, for example, for the searches by \citet{2016AJ....152...47A}, \citet{2016ApJS..222...14V}, \citet{2016A&A...594A.100B}, \citet{2016AJ....152...61M}, \citet{2017MNRAS.464..850G}, \citet{2018MNRAS.474.4603V}, \citet{2018AJ....156...78L}, \citet{2018AJ....155..136M}, and \citet{2018A&A...615L..13G}, or its implementation within the {\tt TERRA} pipeline \citep{2012PASP..124.1073P,2013ApJ...770...69P}, for instance, for the searches by \citet{2015ApJ...804...10C,2016ApJS..226....7C,2018ApJS..239....5C}, \citet{2016ApJ...827...78S}, \citet{2018AJ....156...22Y}, \citet{2018AJ....156..277L}, and \citet{2019RNAAS...3b..43Z}, or similar algorithms searching for box-like flux decreases in stellar light curves \citep{2015ApJ...806..215F} or by eye \citep{2016MNRAS.457.2273O,2016AJ....151..159S}.

The {\tt BLS} algorithm and its variations \citep{2006MNRAS.373..799C,2008A&A...492..617R,2013ApJ...765..132C,2013ApJ...770...69P,2014A&A...561A.138O,2014IAUS..293..410B} is certainly the most widely used transit search algorithm, although alternative algorithms exist \citep{1996Icar..119..244J,2002A&A...395..625A,2004MNRAS.350..331A,2006MNRAS.365..165S,2007ASPC..366..145B,2007A&A...467.1345R,2012A&A...548A..44C}. The popularity of {\tt BLS} is well founded in its good detection efficiency of shallow transits \citep{2003A&A...403..329T,2003A&A...408L...5T} and computational speed, for example, compared to algorithms using artificial intelligence \citep{2016MNRAS.455..626M,2018MNRAS.474..478P,2018AJ....155..147Z,2018MNRAS.478.4225A}.


We recently presented the {\tt transit least-squares} ({\tt TLS}) algorithm, which is optimized for the detection of shallow periodic transits \citep{2019A&A...623A..39H}. In contrast to {\tt BLS}, the test function of {\tt TLS} is not a box, but an analytical model of a transit light curve \citep{2002ApJ...580L.171M}. As a consequence, the residuals between the {\tt TLS} search function and the observed data are substantially smaller than the residuals obtained with {\tt BLS} or similar box-like algorithms, resulting in an enhancement of the signal detection efficiency for {\tt TLS}, in particular for weak signals. 

The advantages of using {\tt TLS} instead of {\tt BLS} in searching for small planets were first illustrated by \citet{2019A&A...623A..39H} using simulated light curves. These predictions were verified using actual {\it K2} photometry with the discovery of the Earth-sized planet K2-32\,e \citep[][hereafter \PaperI]{2019A&A...625A..31H} around a star that was known to host three roughly Neptune-sized planets. Here we extend the {\tt TLS} Survey and present the results of our search for hitherto unknown additional planets around all stars from {\it K2} that were already known to host at least one planet.

\section{Methods}
\label{sec:methods}

\subsection{Target selection}

We reanalyzed the 517 {\it K2} planets and candidates (in 489 unique systems) from the NASA Exoplanet Archive\footnote{\href{https://exoplanetarchive.ipac.caltech.edu}{https://exoplanetarchive.ipac.caltech.edu}} \citep{2013PASP..125..989A} and the Exoplanet Orbit Database\footnote{\href{http://exoplanets.org}{http://exoplanets.org}} \citep{2014PASP..126..827H}, most of which have been discovered by \citet{2015ApJ...806..215F}, \citet{2016ApJS..222...14V}, \citet{2016ApJS..226....7C}, \citet{2018AJ....155..136M}, and \citet{2018AJ....156...78L}. We chose these {\it K2} systems with known planets for this initial phase of the {\tt TLS} Survey since any additional findings would have an intrinsically low false-positive probability \citep{2012ApJ...750..112L}, which simplifies our vetting and validation procedure.

\subsection{Transit search}

Our methods are in principle the same as described in \PaperI. We used the publicly available {\it K2} light curves after correction for instrumental effects with {\tt EVEREST} \citep{2016AJ....152..100L}. Using the published periods, transit durations, and mid-transit times, we removed the in-transit flux of known planets in these systems. Then we removed stellar variability and other trends using a median filter with a window size of 25 long cadences. Each cadence corresponds to an exposure of about 30\,min, making the walking window as wide as about half a day. This window size was chosen as a compromise so that it is sufficiently short to remove stellar variability, but leaves transit signals intact. For planets with periods $<80\,$days, transit durations are shorter in any physically plausible case, and much shorter in almost all cases (\citealt{2019A&A...623A..39H}, \PaperI). The longest transit duration of the planets we discover is about 0.21\,d. We clipped all data points that are more than $3\,\sigma$ above the running mean in order to remove data points affected by cosmic-ray impacts on the CCD.

After this preprocessing of the light curves, we applied {\tt TLS} (version 1.0.13) using the stellar limb darkening, mass, and radius estimates available in the EPIC catalog \citep{2016ApJS..224....2H}. For all other {\tt TLS} parameters we used the default values.

\subsection{Vetting}
\label{sec:vetting}

\begin{figure*}[th!]
\centering
\includegraphics[width=.82\linewidth]{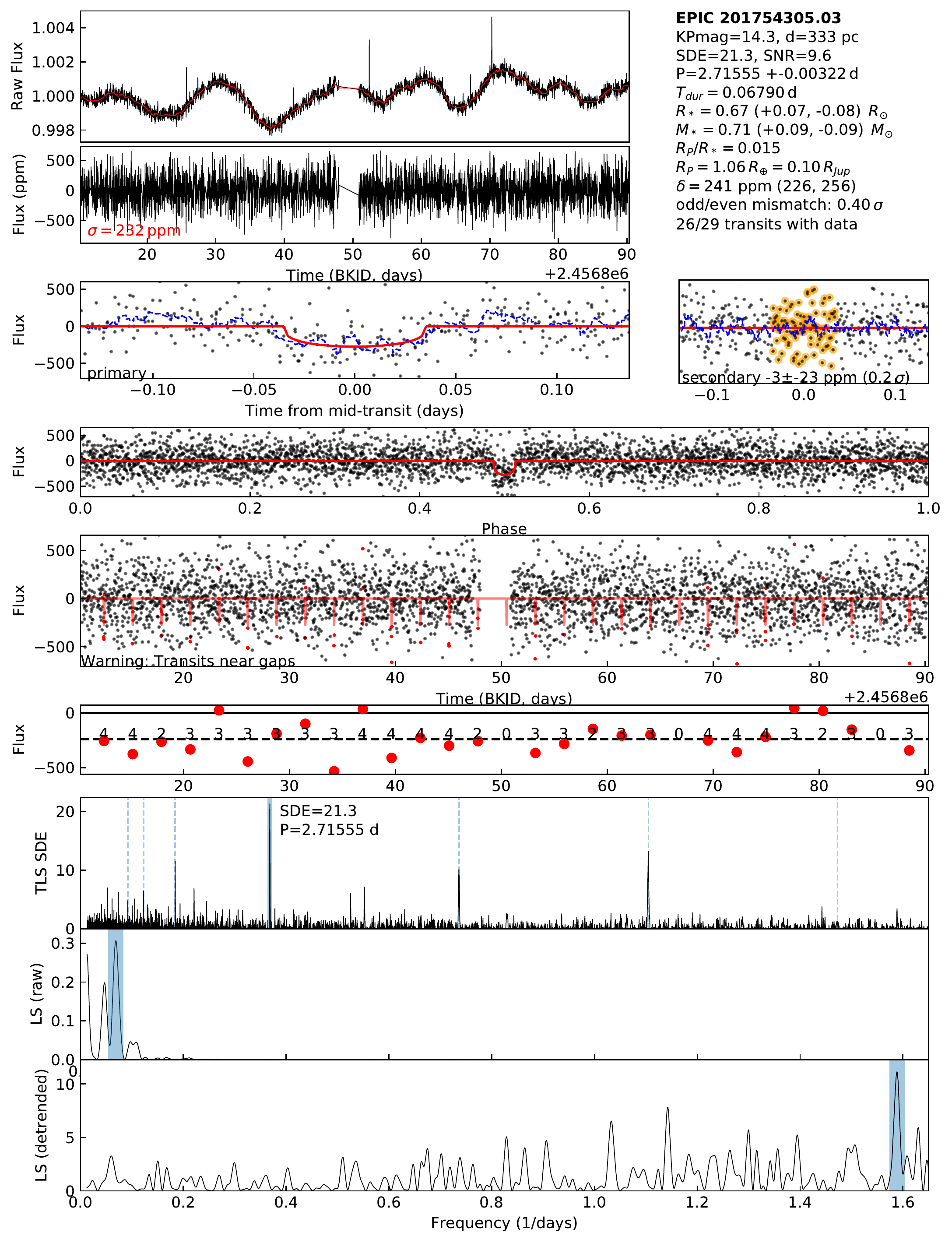}
\caption{Example of a vetting sheet, here for 201754305.03. From top to bottom, the panels show the raw flux of the {\tt EVEREST} light curve (with the running median in red), the normalized light curve, zooms into the transit candidate (left), and possible eclipse (right) times of the phase-folded light curve, the entire phase-folded light curve, the entire light curve with in-transit data of the new candidate marked in red, the sequence of transit depth measurements (including the number of in-transit points), the SDE$_{\rm TLS}$ periodogram, the Lomb-Scargle periodogram of the raw light curve, and the Lomb-Scargle periodogram of the detrended light curve. A summary of the basic system properties is shown in the upper right corner. 
}
\label{fig:vetting}
\end{figure*}

\begin{figure*}
\centering
\includegraphics[width=.420\linewidth]{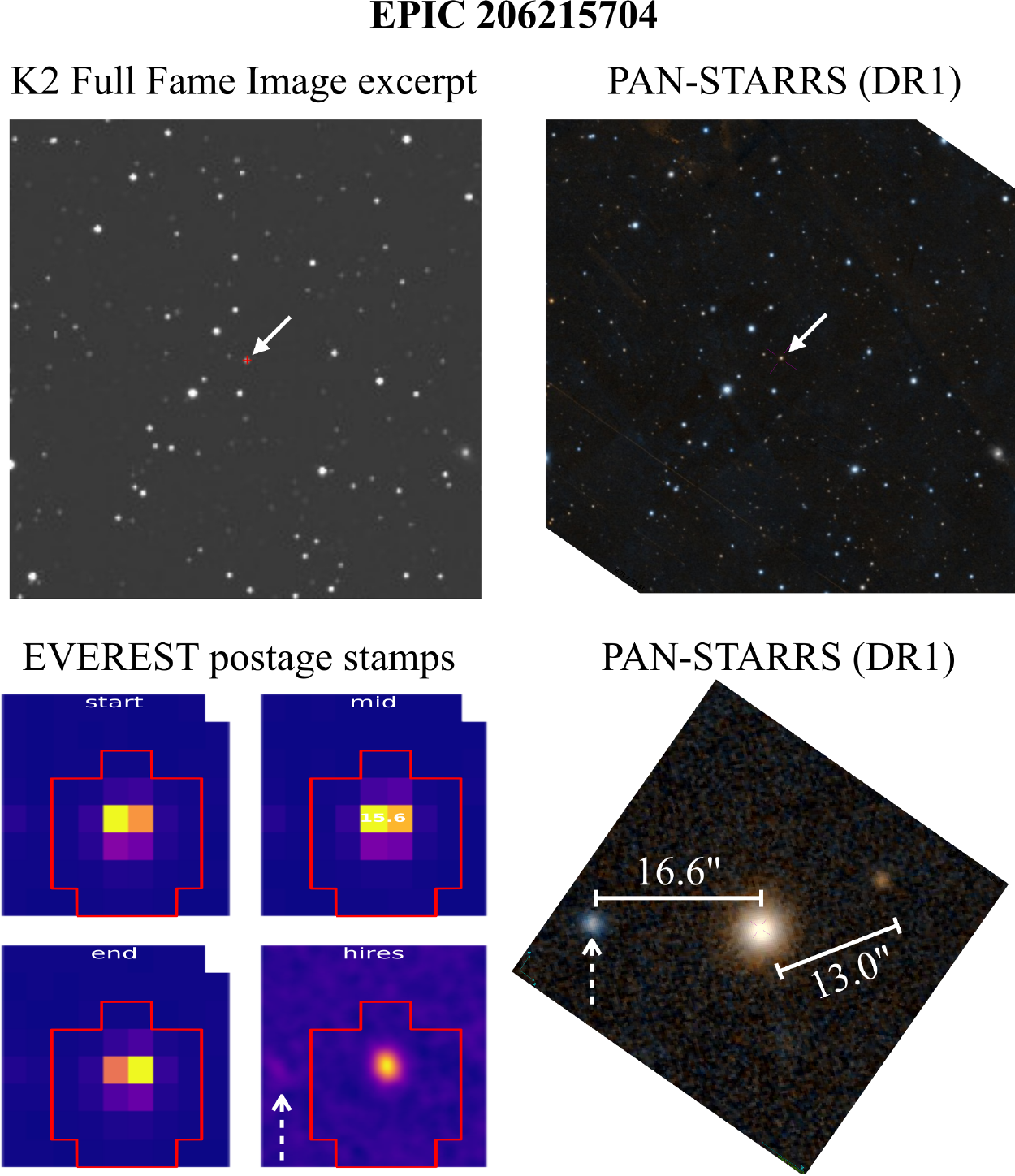}
\hspace{1.4cm}
\includegraphics[width=.420\linewidth]{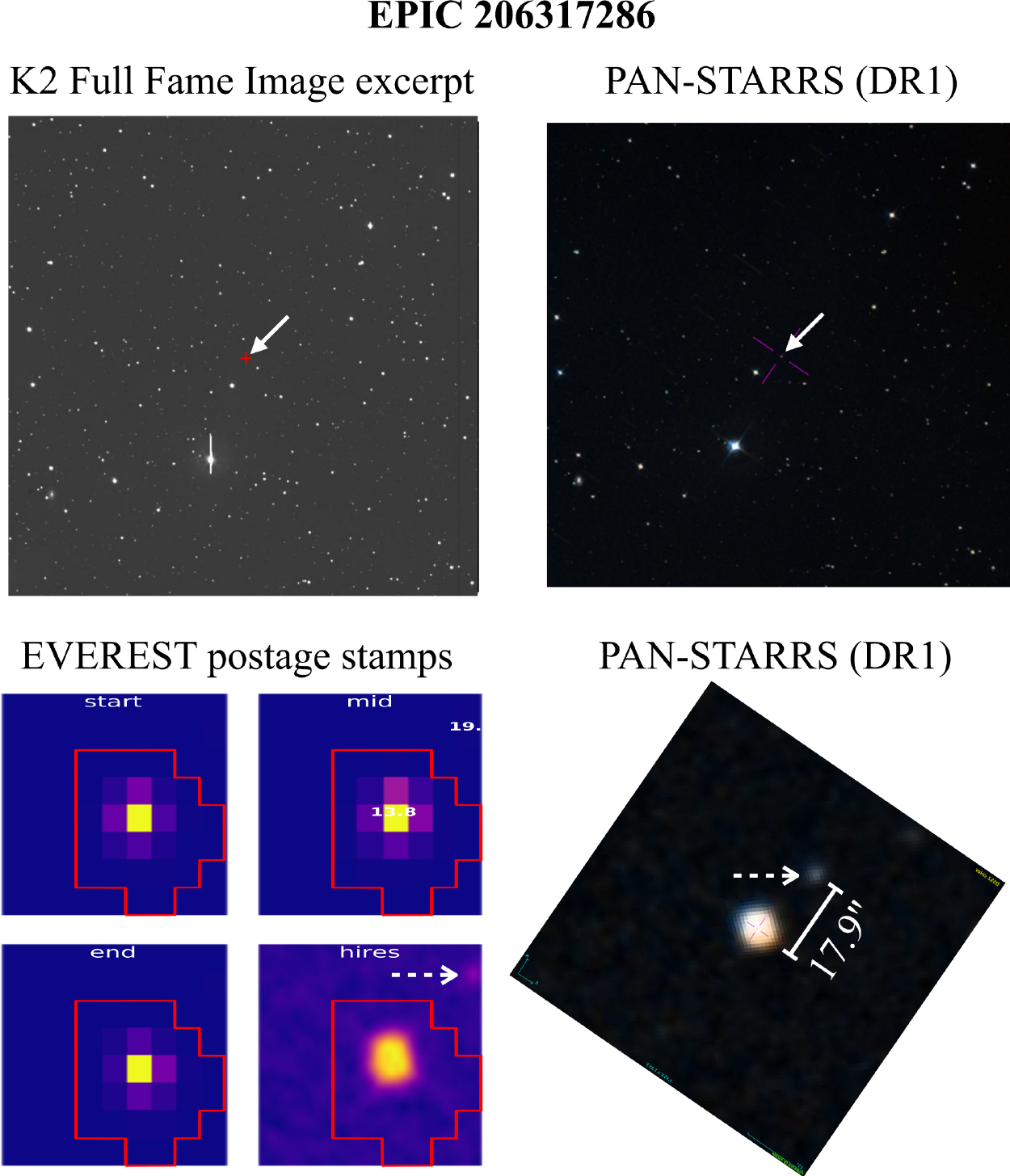}\\
\vspace{0.6cm}
\includegraphics[width=.410\linewidth]{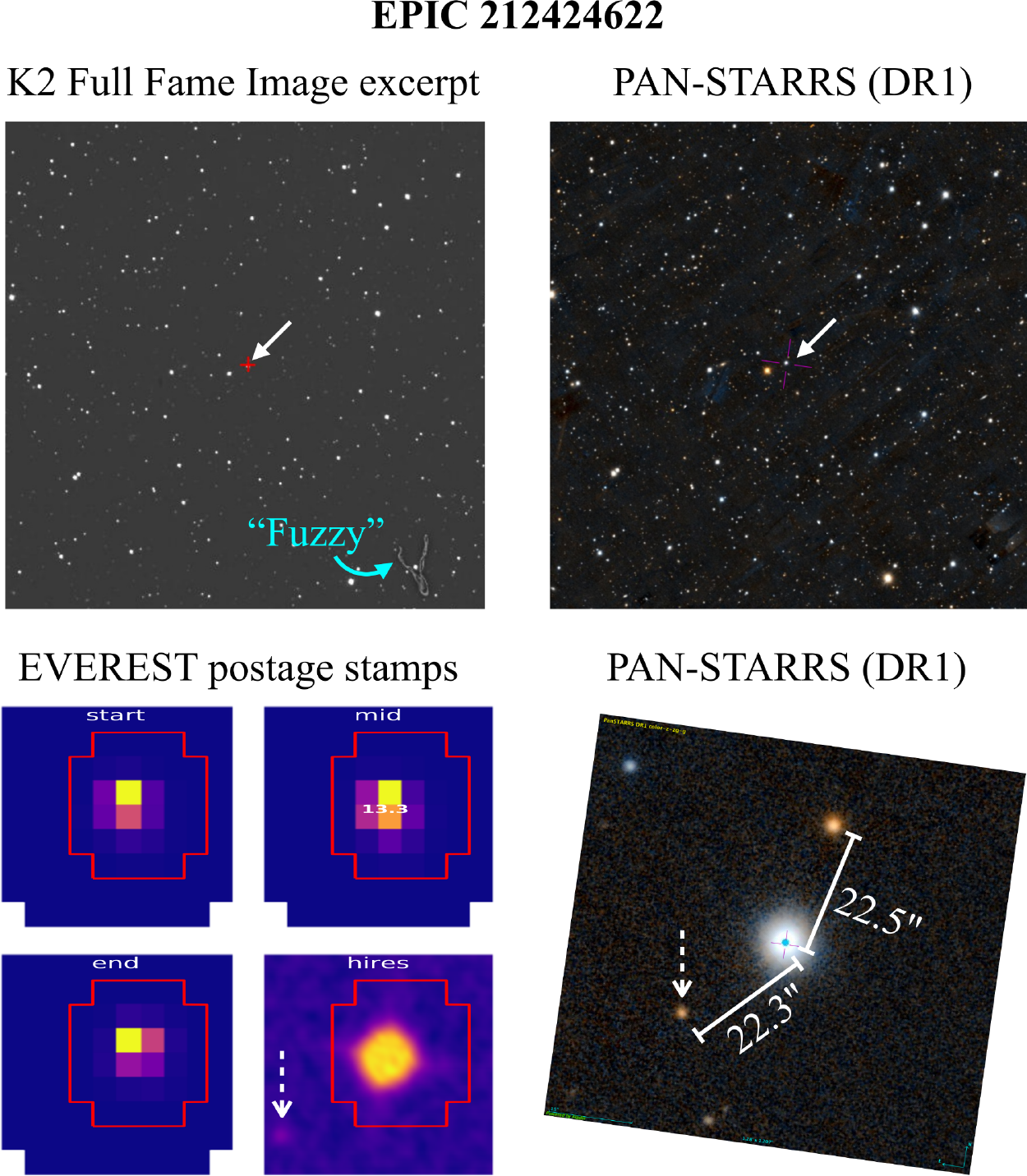}
\hspace{1.4cm}
\includegraphics[width=.420\linewidth]{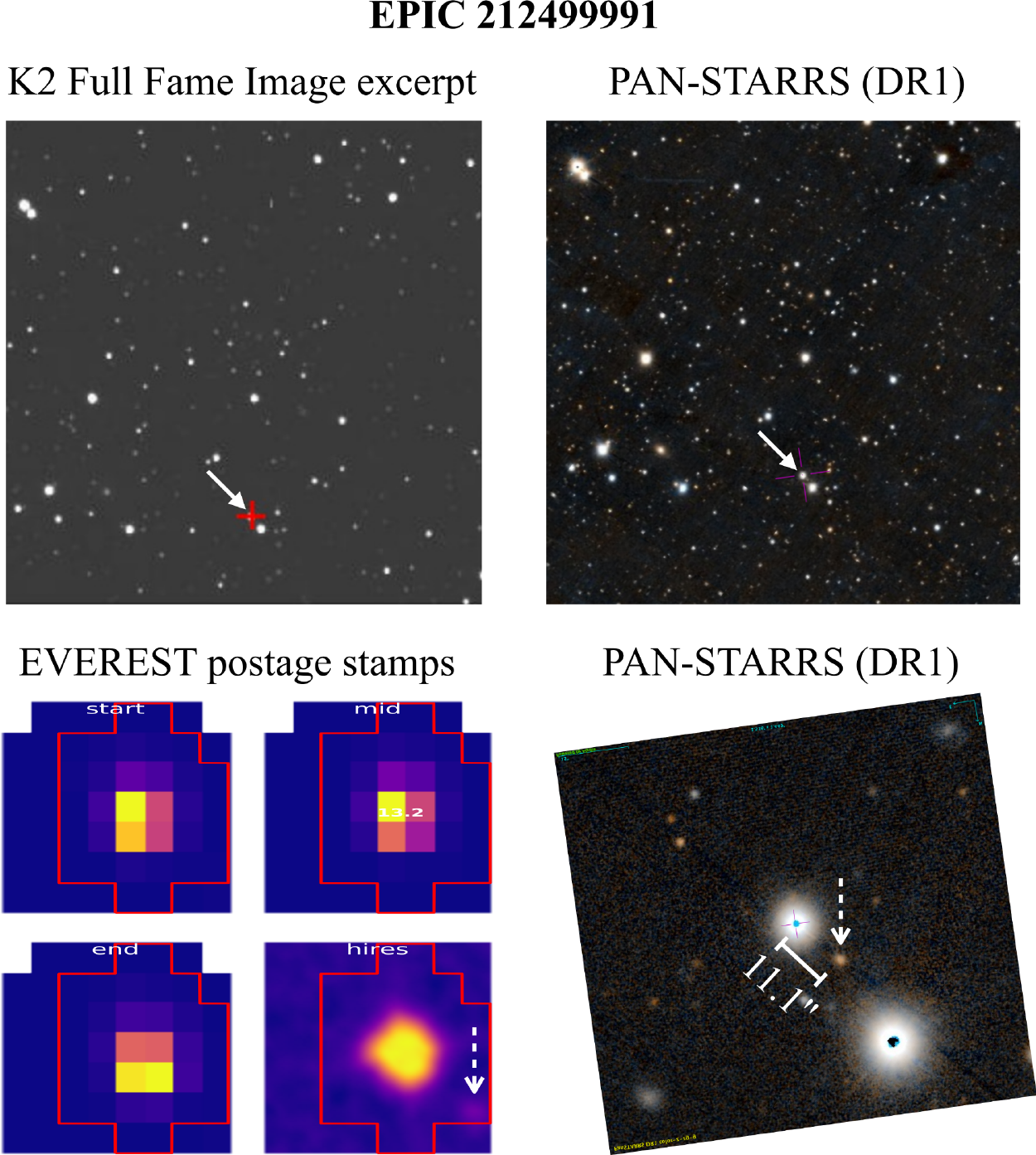}
\caption{Comparison of {\it K2} CCD Full Frame Images, {\it K2} aperture masks used by {\tt EVEREST}, and high-resolution Pan-STARRS (DR1) images for EPIC\,206215704 (top left), EPIC\,206317286 (top right), EPIC\,212424622 (bottom left), and EPIC\,212499991 (bottom right). For each of the four objects, four panels show in a clockwise direction the {\it K2} CCD Full Frame Image, the Pan-STARRS image, a Pan-STARRS image zoom, and the {\tt EVEREST} postage stamps with the pixel map of the aperture mask outlined in red. Solid arrows mark the positions of our targets. Dashed arrows indicate nearby sources that we examined as possible contaminants. ``Fuzzy'' is labeled and marked with a curved light blue arrow.}
\label{fig:PANSTARRS_contaminants}
\end{figure*}

\begin{figure*}[t]
\centering
\includegraphics[width=0.9\linewidth]{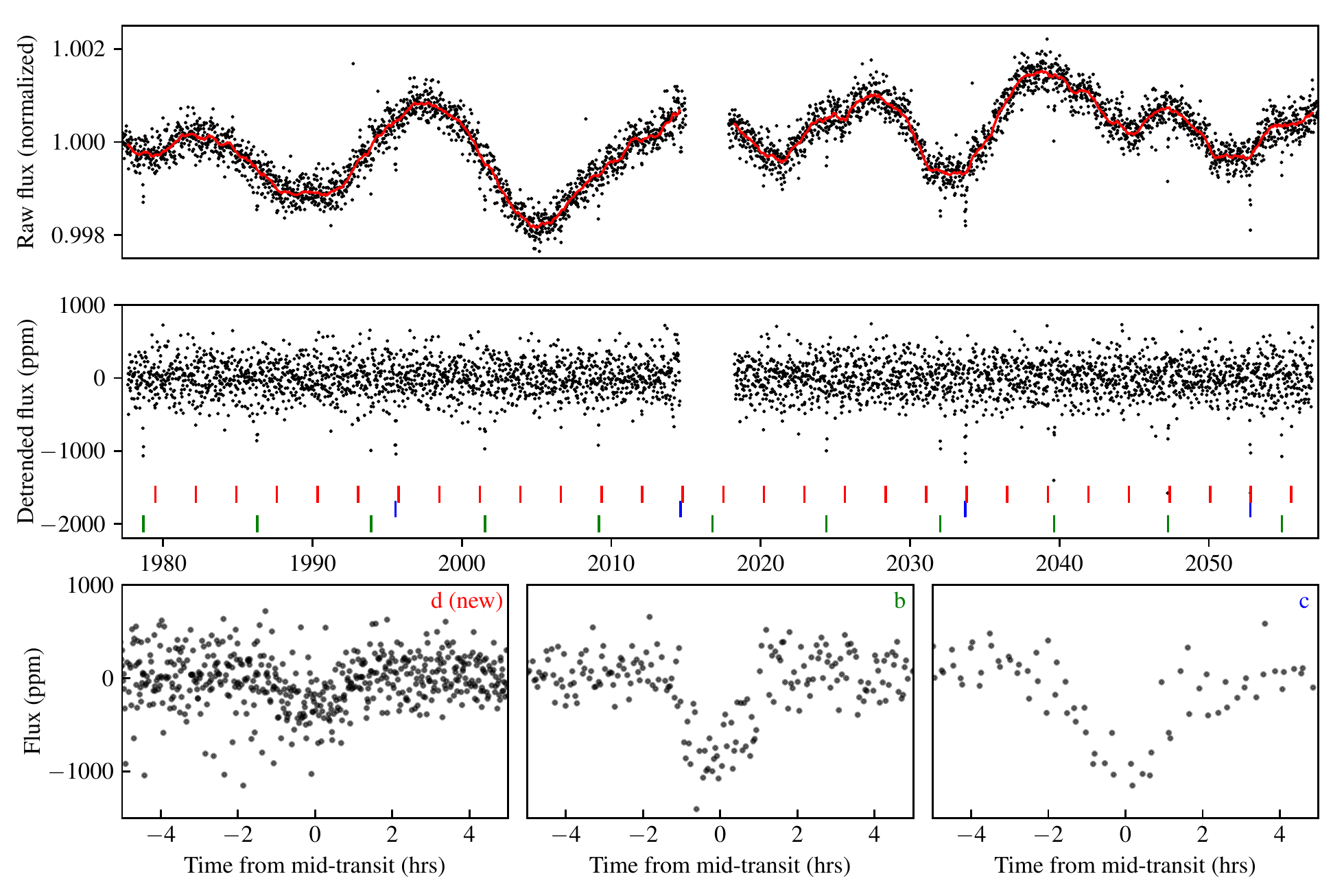}
\caption{Example of our detrending of a {\it K2} light curve and successive transit detection with {\tt TLS} using the data of K2-16 (EPIC\,201754305). {\it Top}: {\tt EVEREST} light curve (black dots) and our running median filter (red line). {\it Center}: Detrended light curve that we used as an input to {\tt TLS}. Transit times of the previously known planets K2-16\,b and c are marked with green and blue vertical bars, respectively. Transit times of the newly discovered planet K2-16\,d are indicated with red vertical bars. {\it Bottom}: Phase-folded transit light curves of all three planets sorted by increasing orbital period from left to right: K2-16\,d, b, and c.}
\label{fig:EPIC201754305}
\end{figure*}

\begin{figure*}[h!]
\centering
\includegraphics[width=0.258\linewidth]{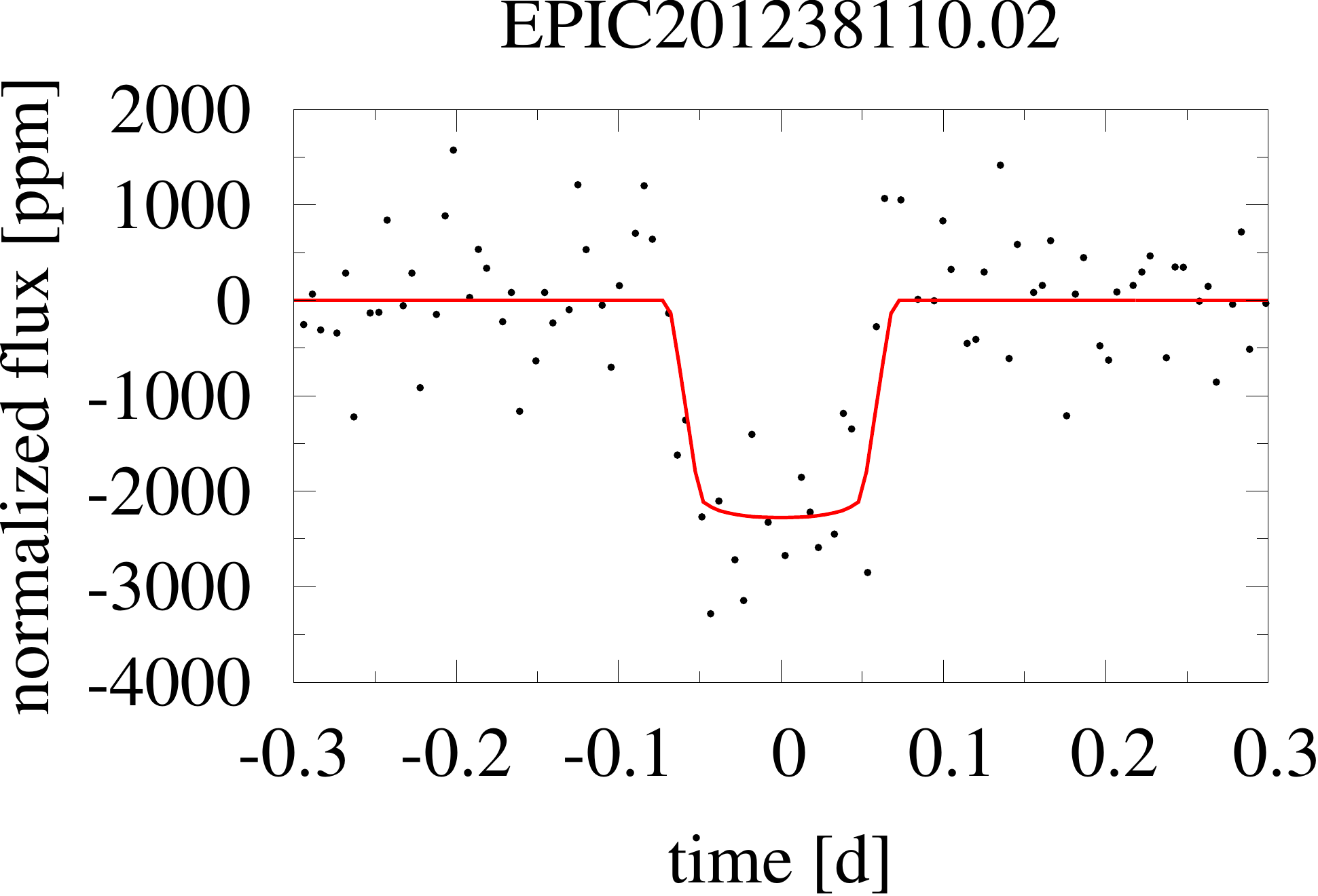}
\includegraphics[width=0.242\linewidth]{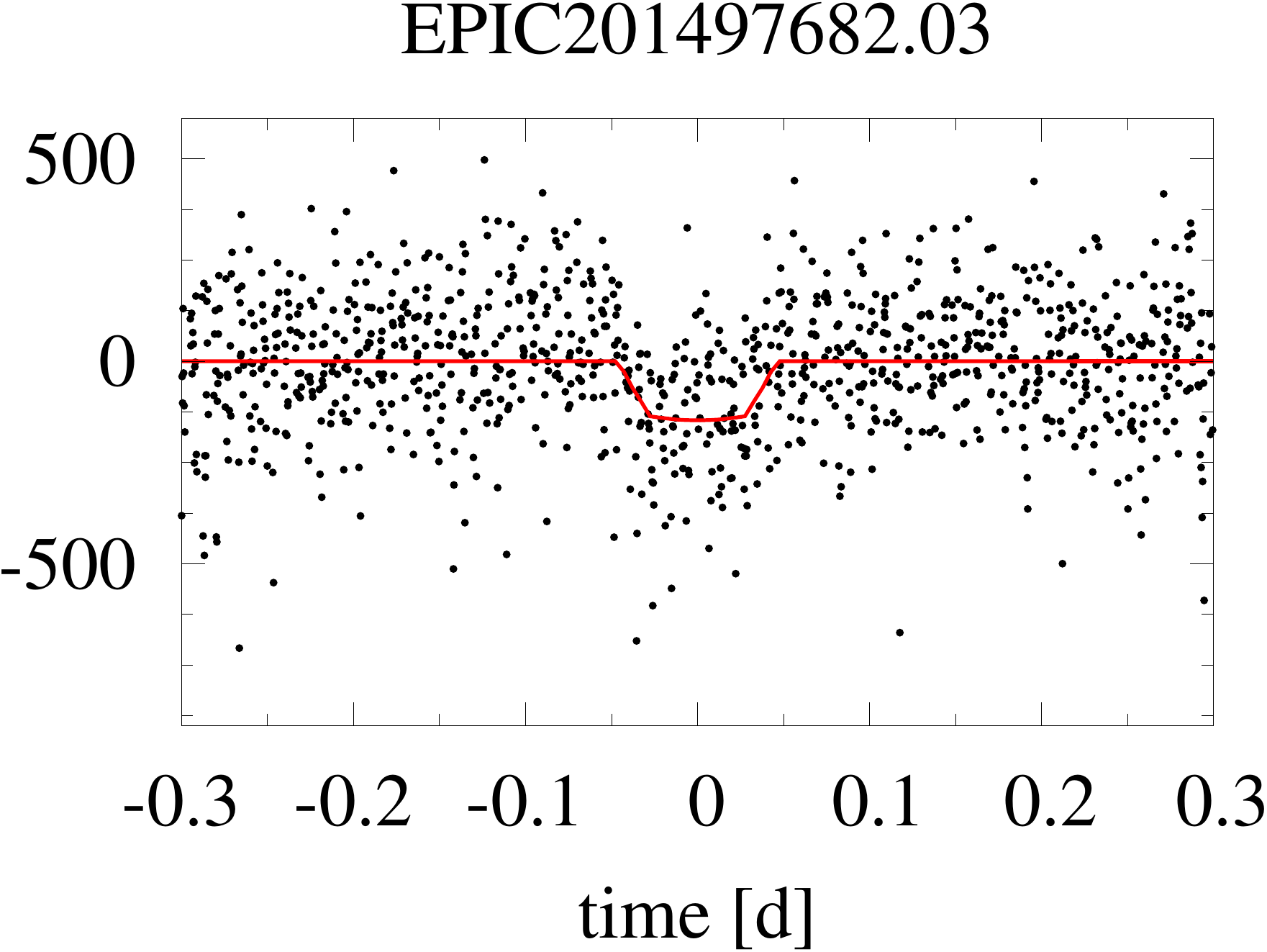}
\includegraphics[width=0.242\linewidth]{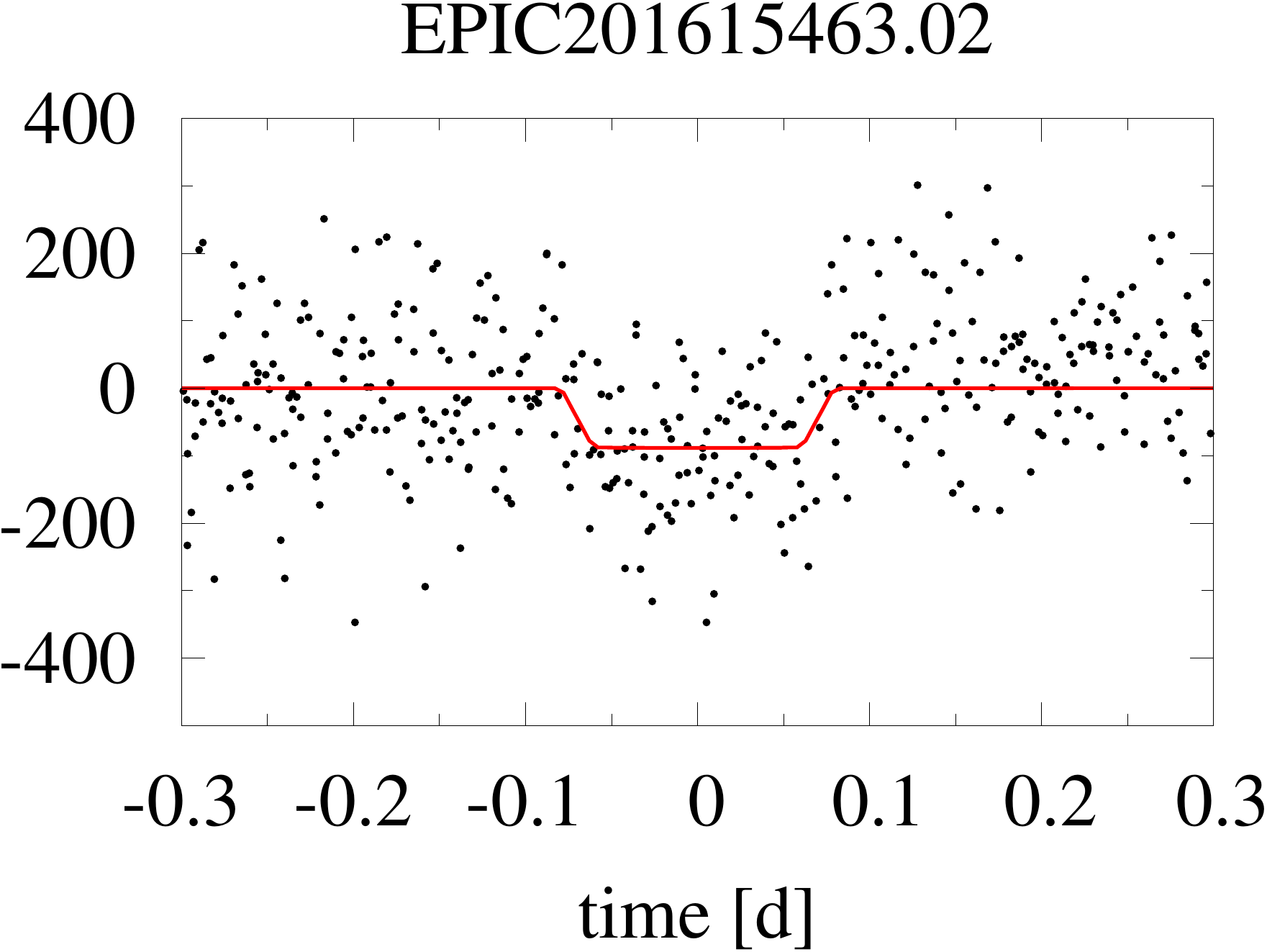}
\includegraphics[width=0.242\linewidth]{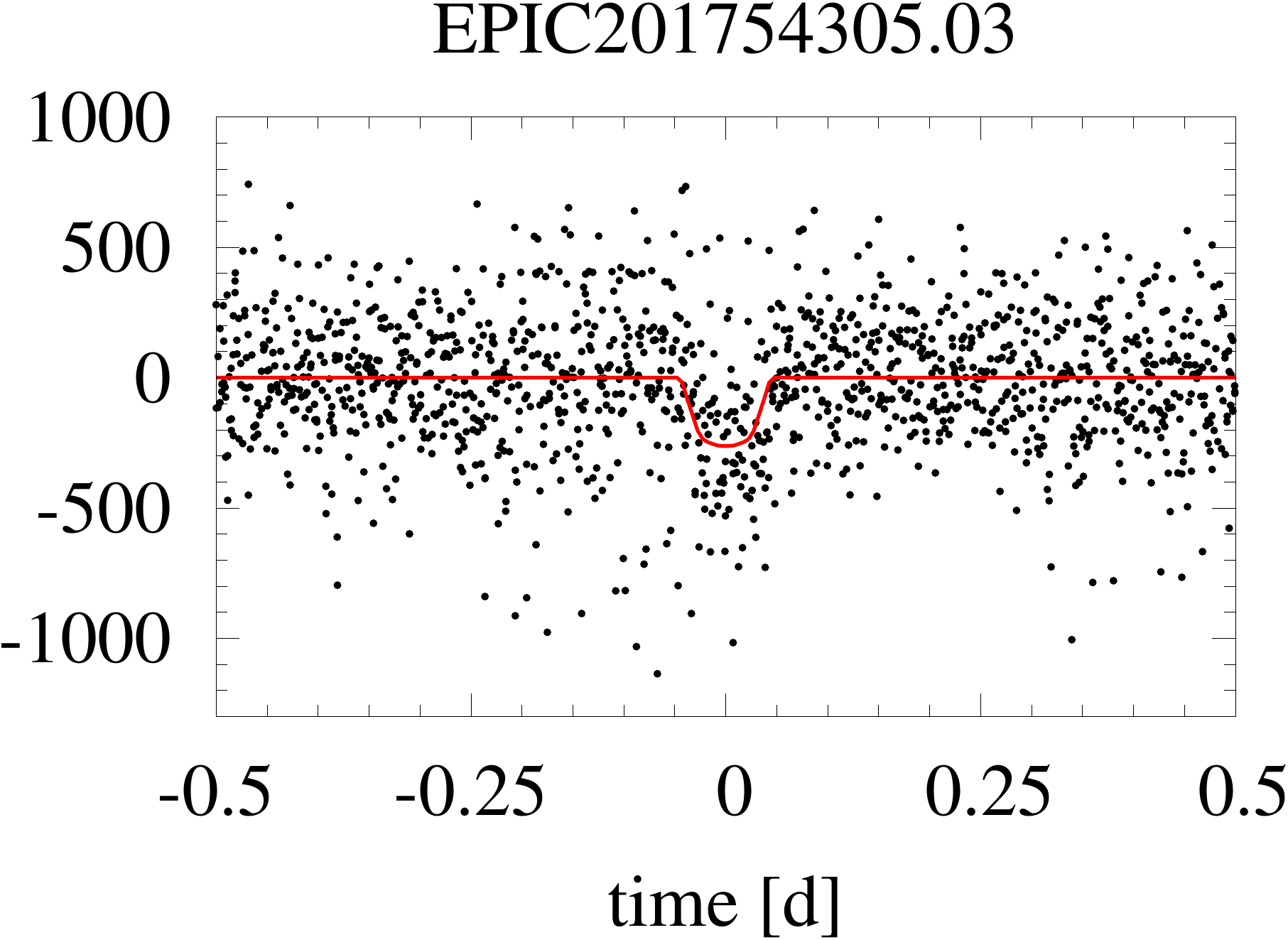}\\
\vspace{.35cm}
\includegraphics[width=0.258\linewidth]{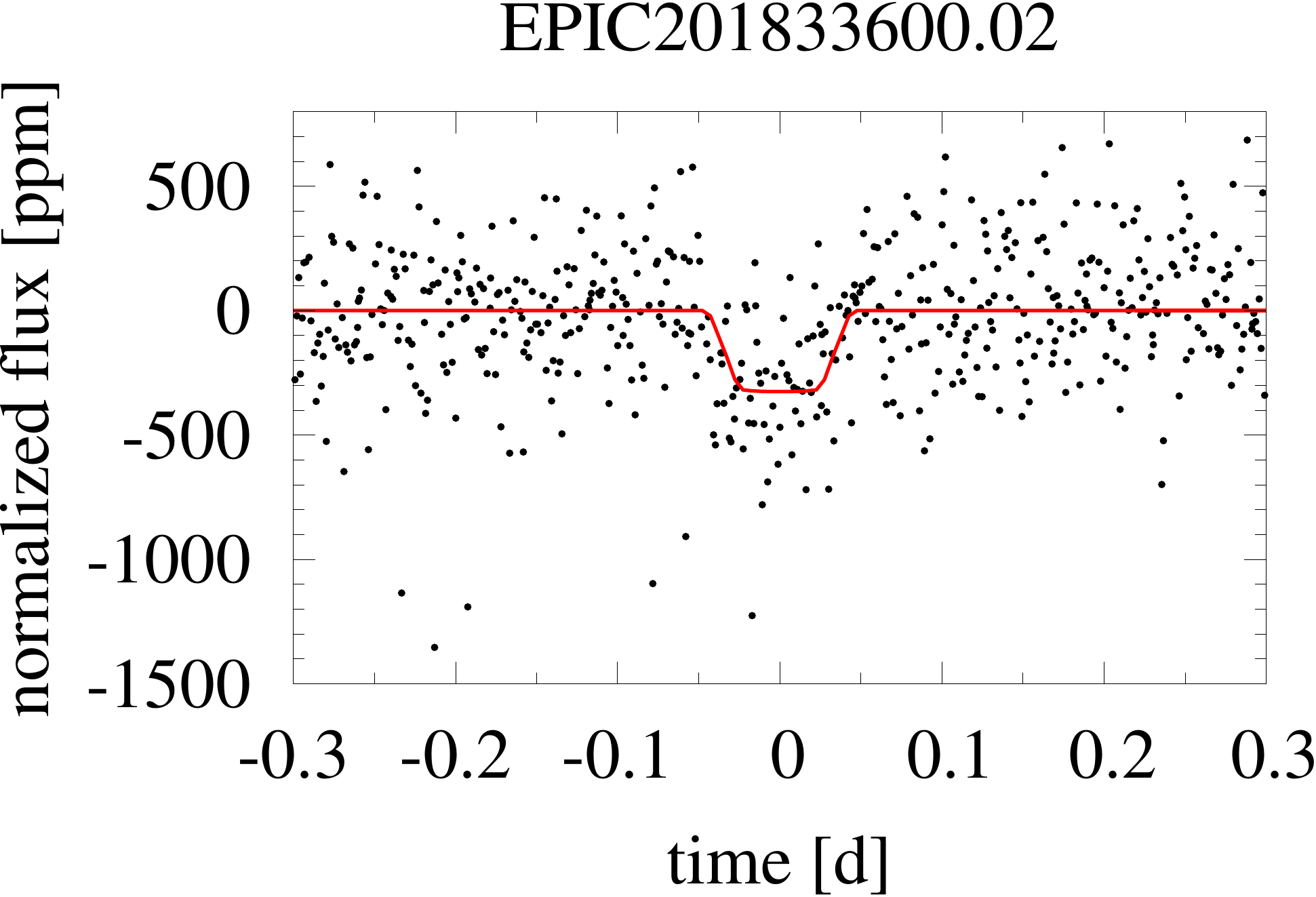}
\includegraphics[width=0.242\linewidth]{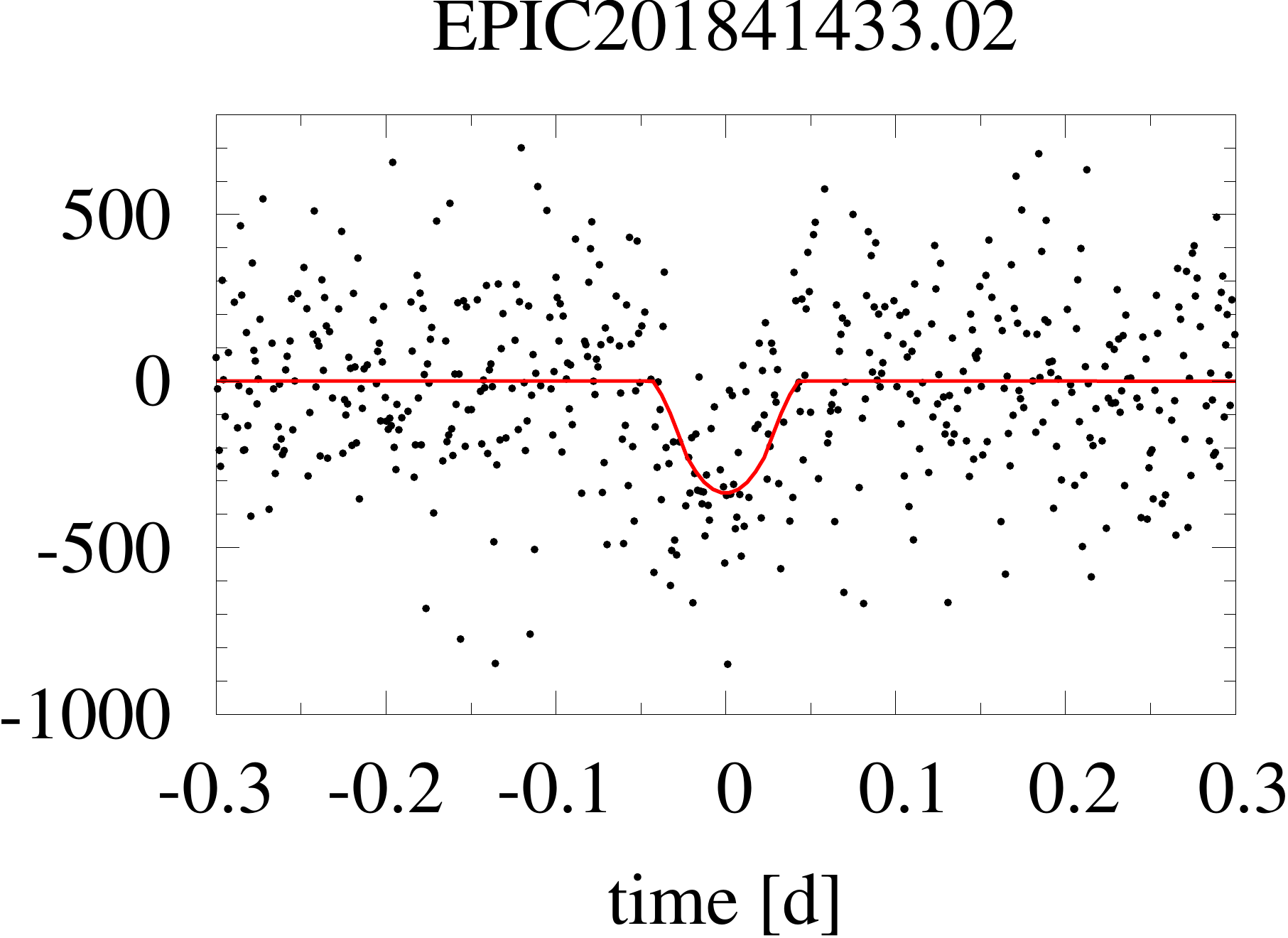}
\includegraphics[width=0.242\linewidth]{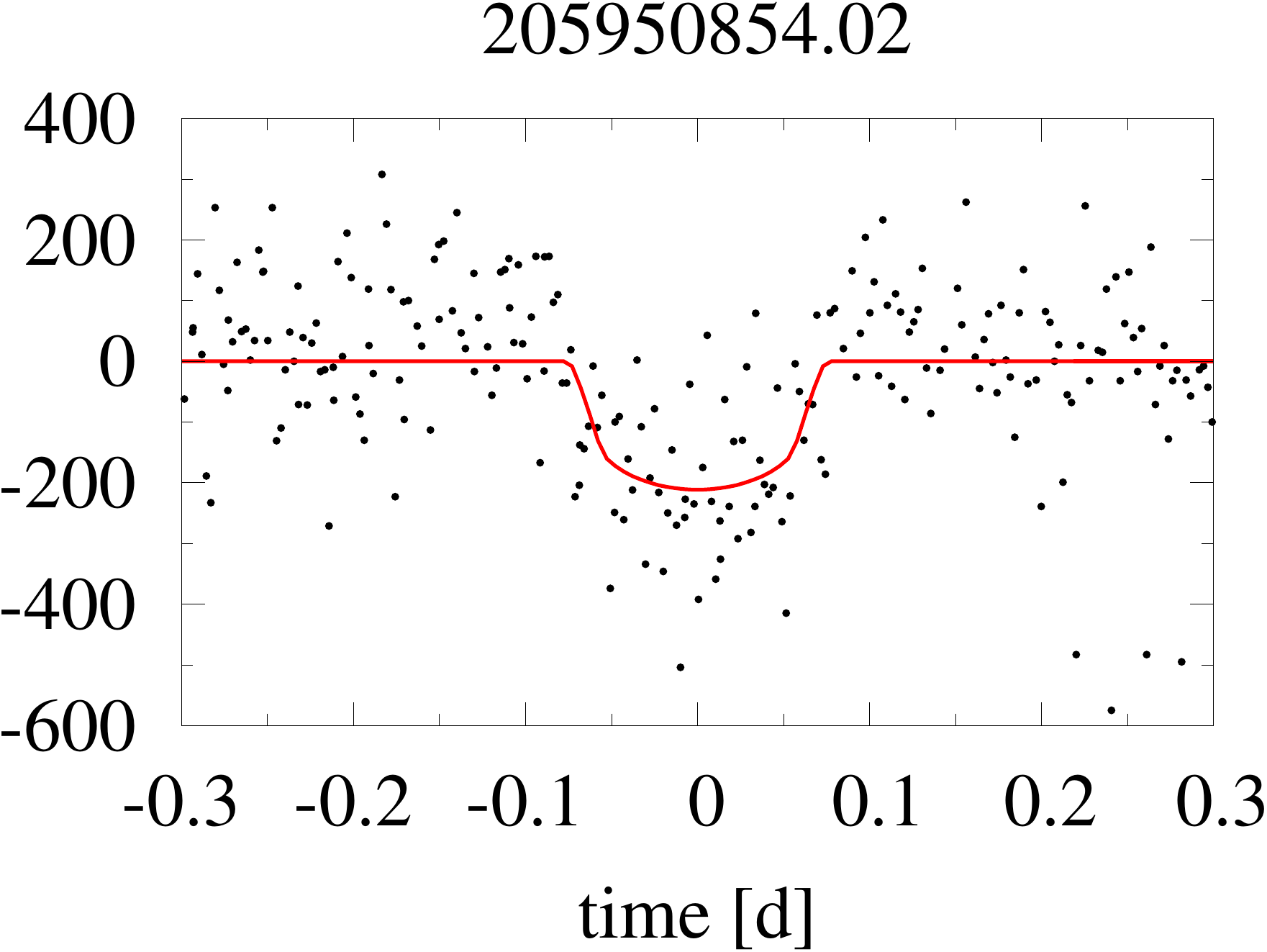}
\includegraphics[width=0.242\linewidth]{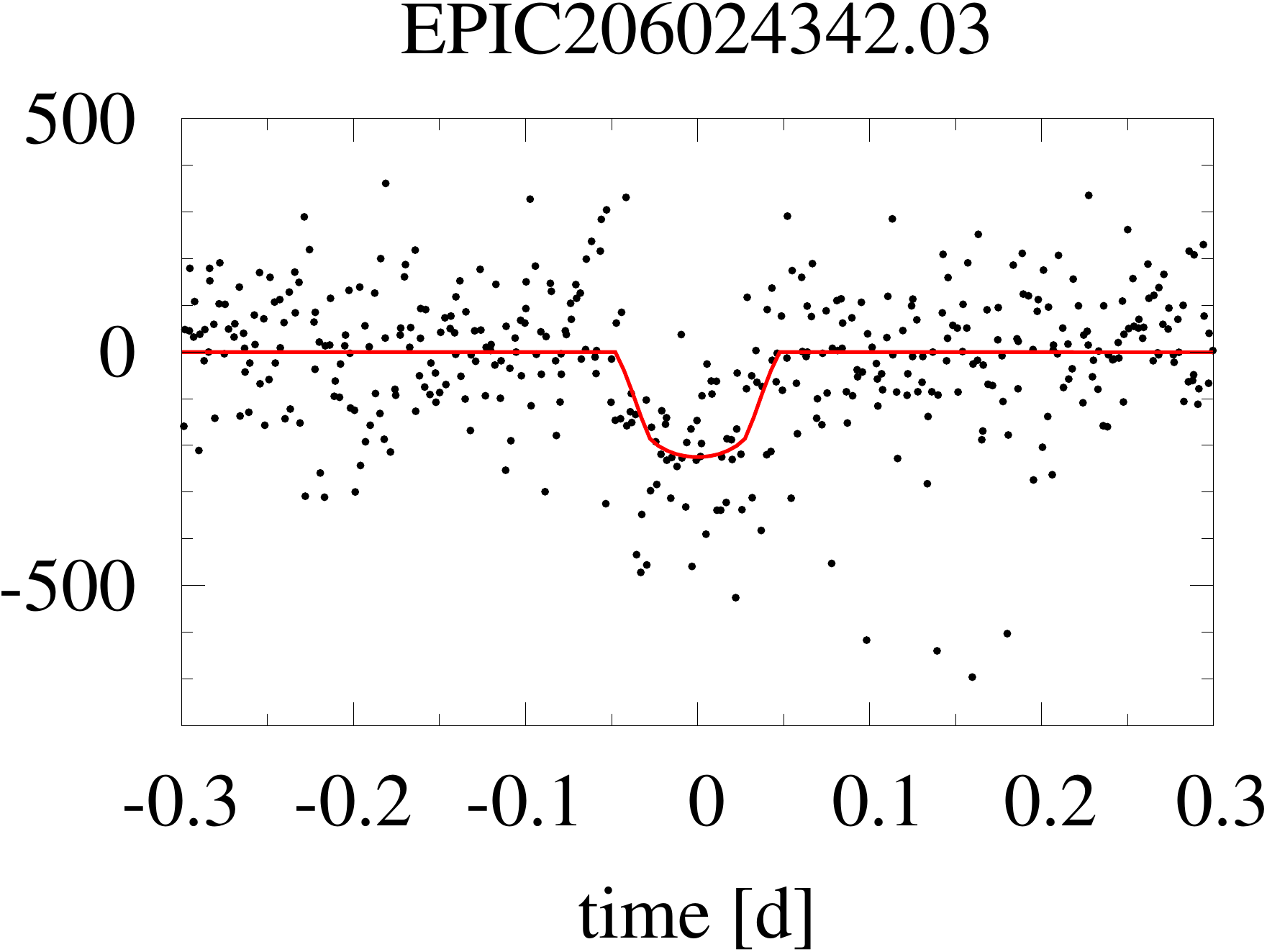}\\
\vspace{.35cm}
\includegraphics[width=0.258\linewidth]{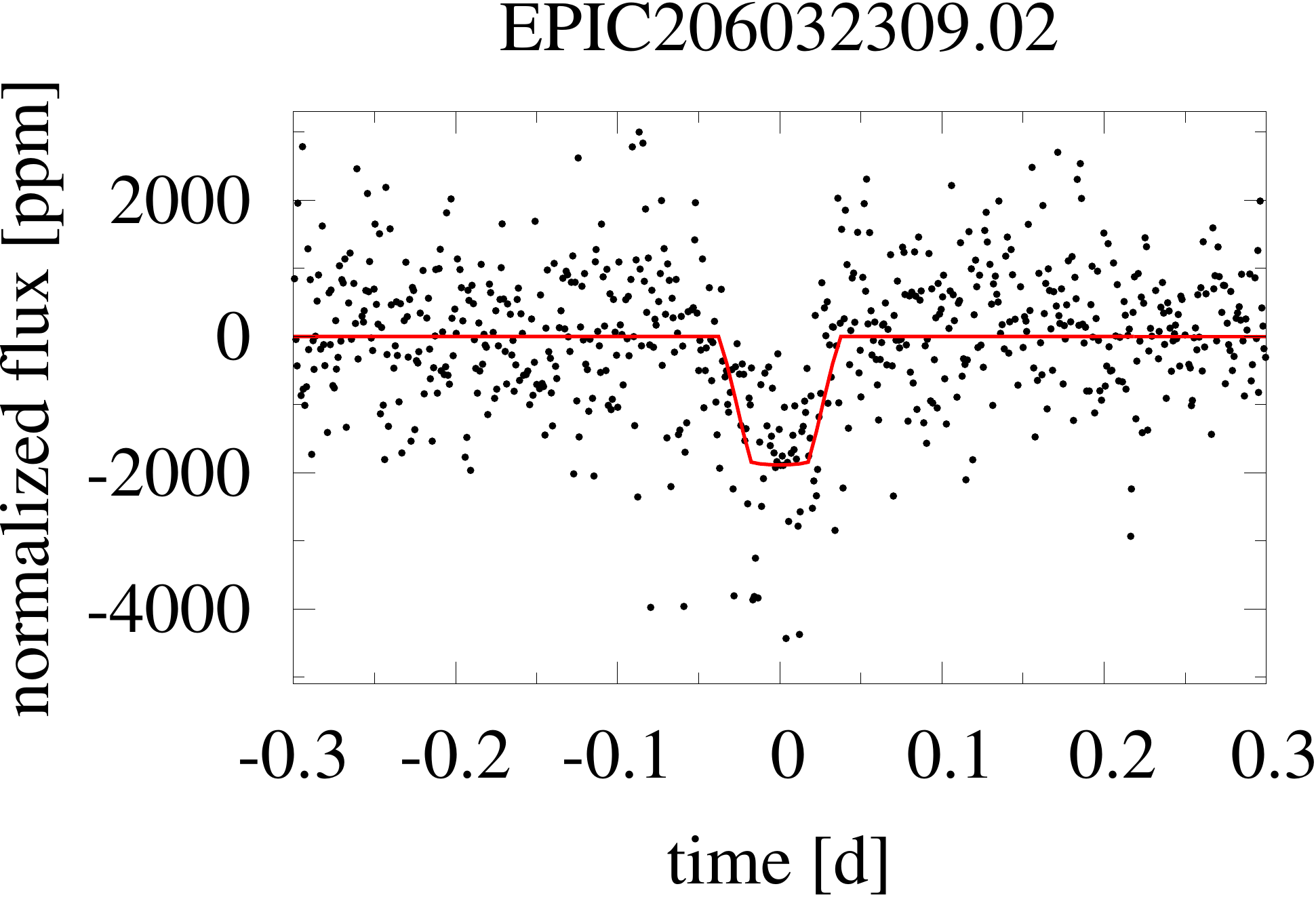}
\includegraphics[width=0.242\linewidth]{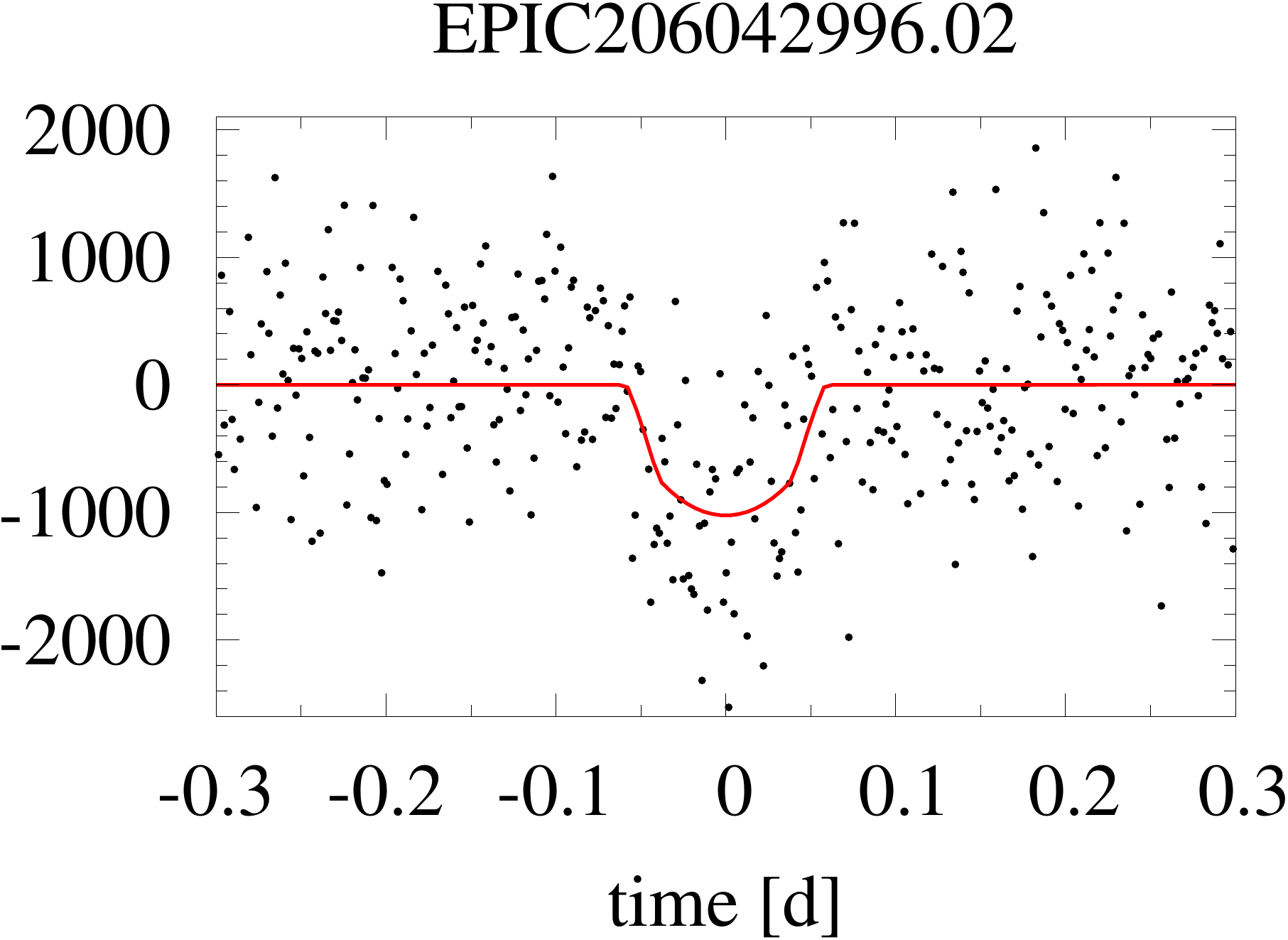}
\includegraphics[width=0.242\linewidth]{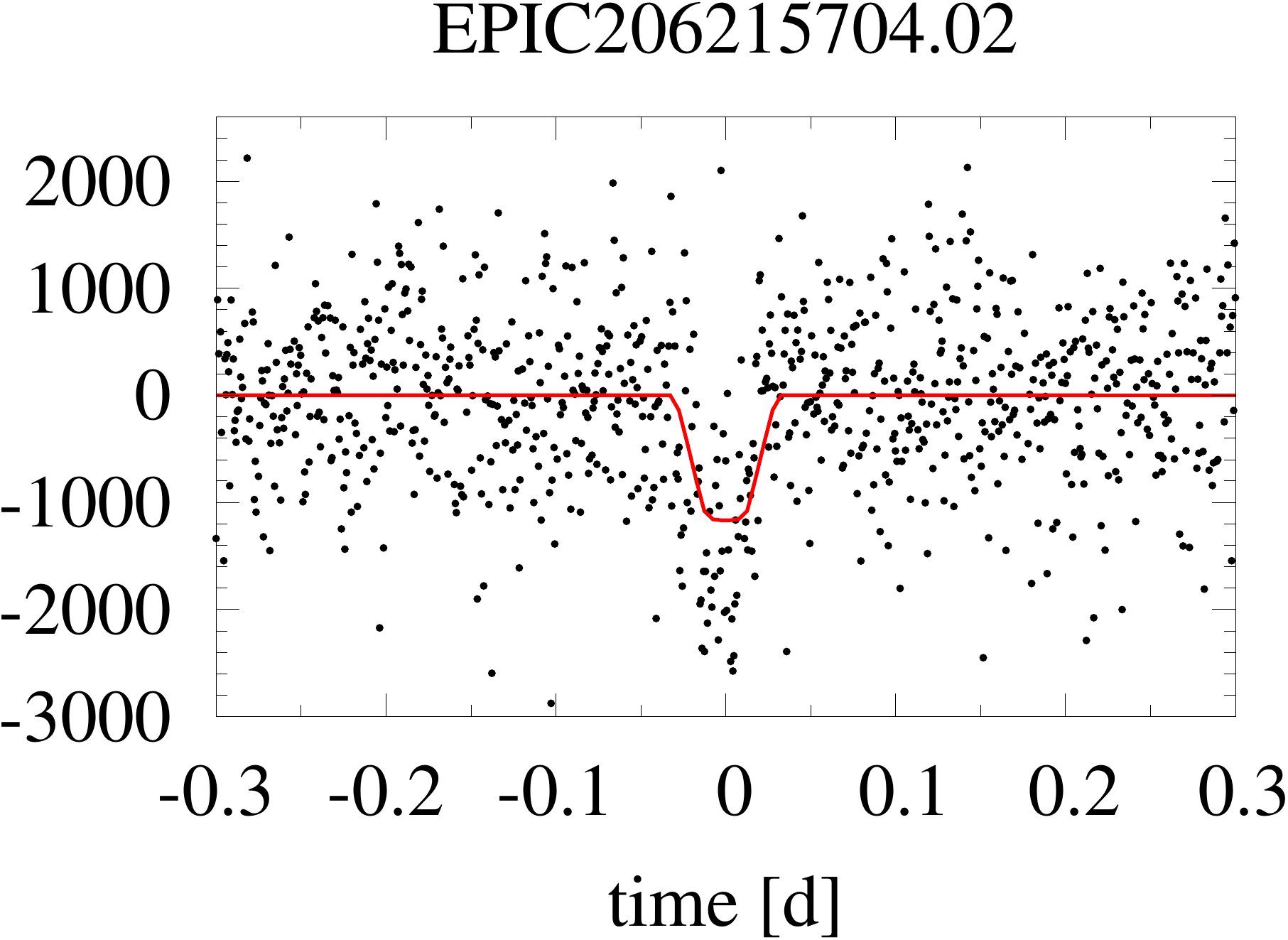}
\includegraphics[width=0.242\linewidth]{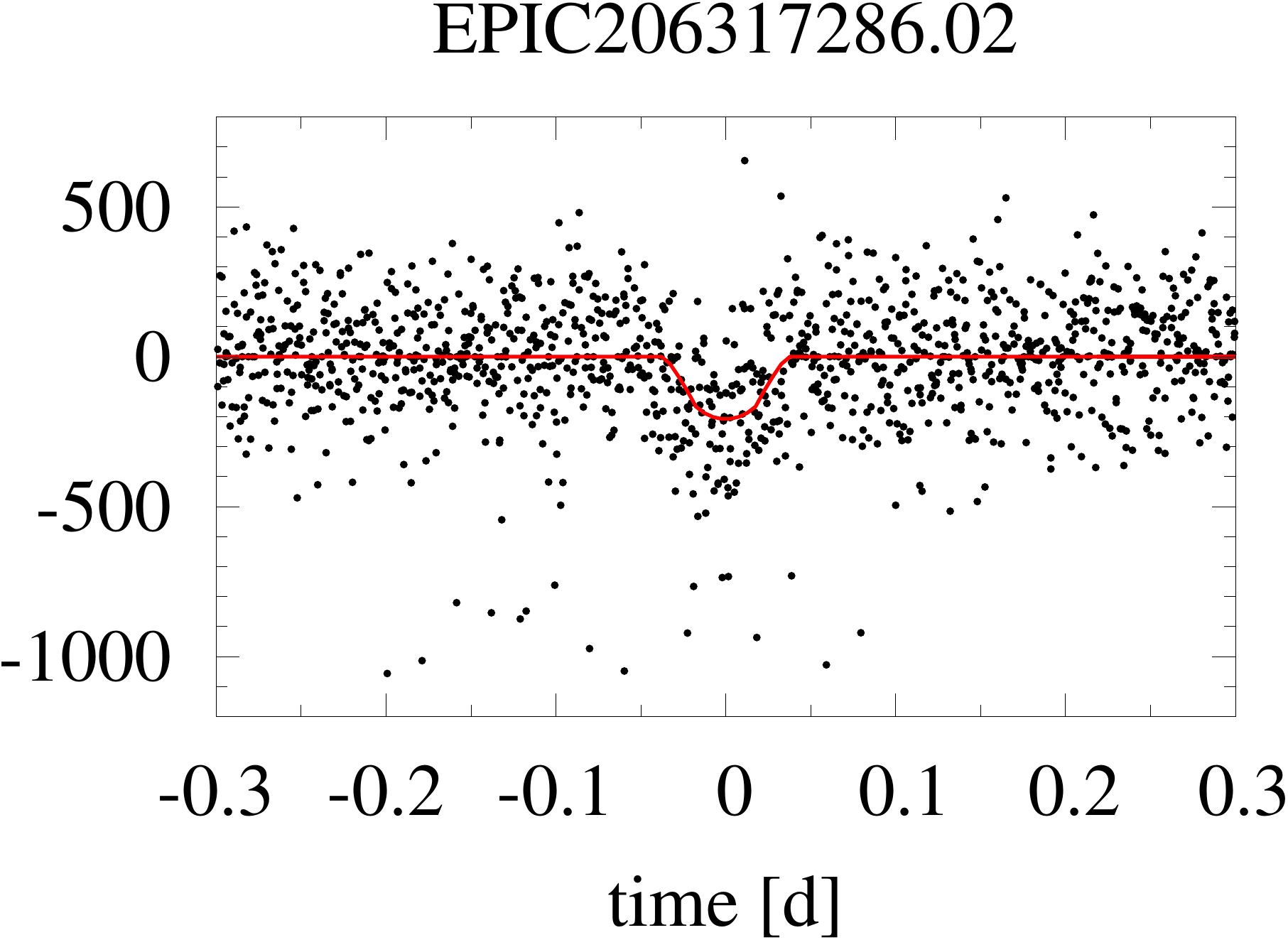}\\
\vspace{.35cm}
\includegraphics[width=0.258\linewidth]{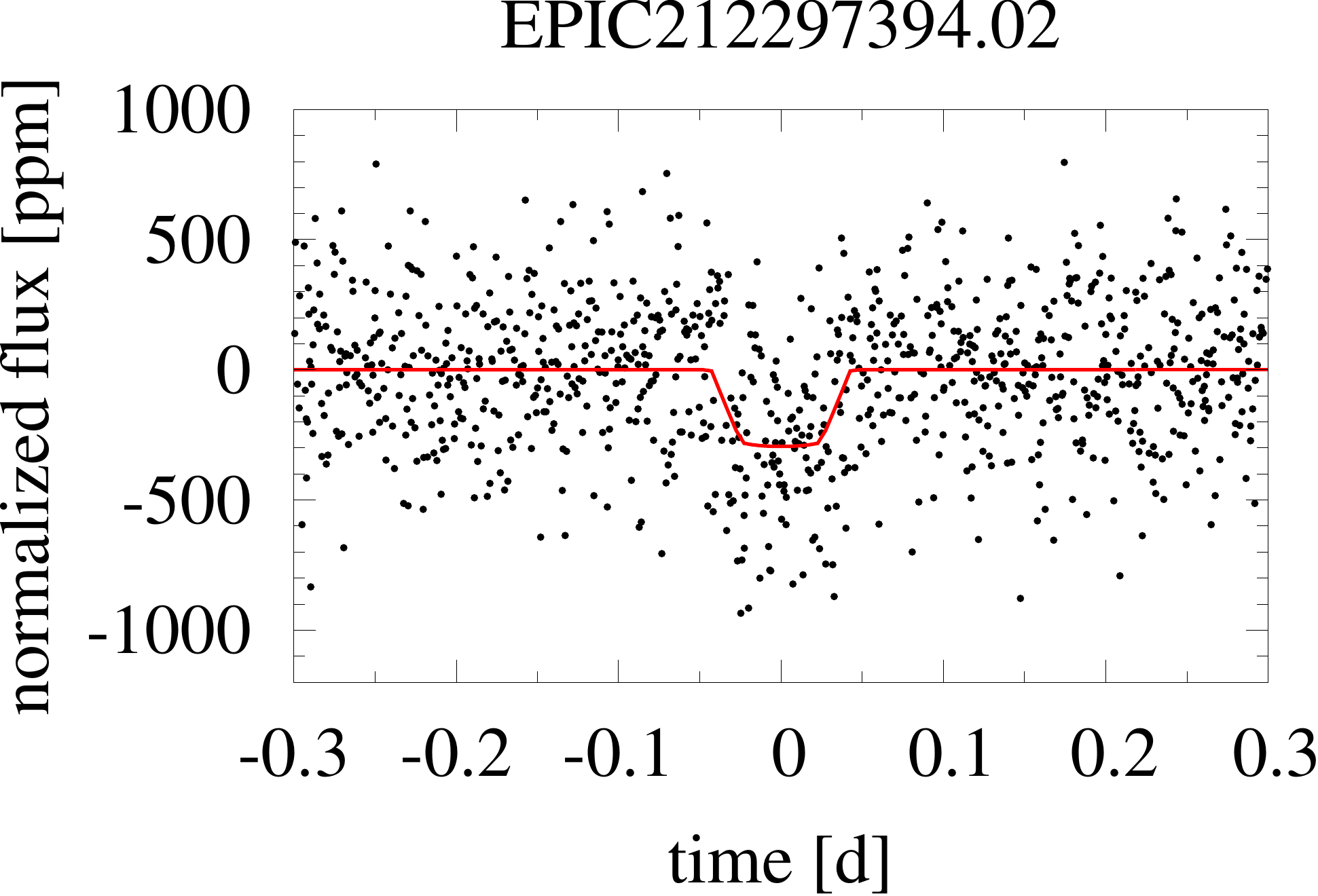}
\includegraphics[width=0.242\linewidth]{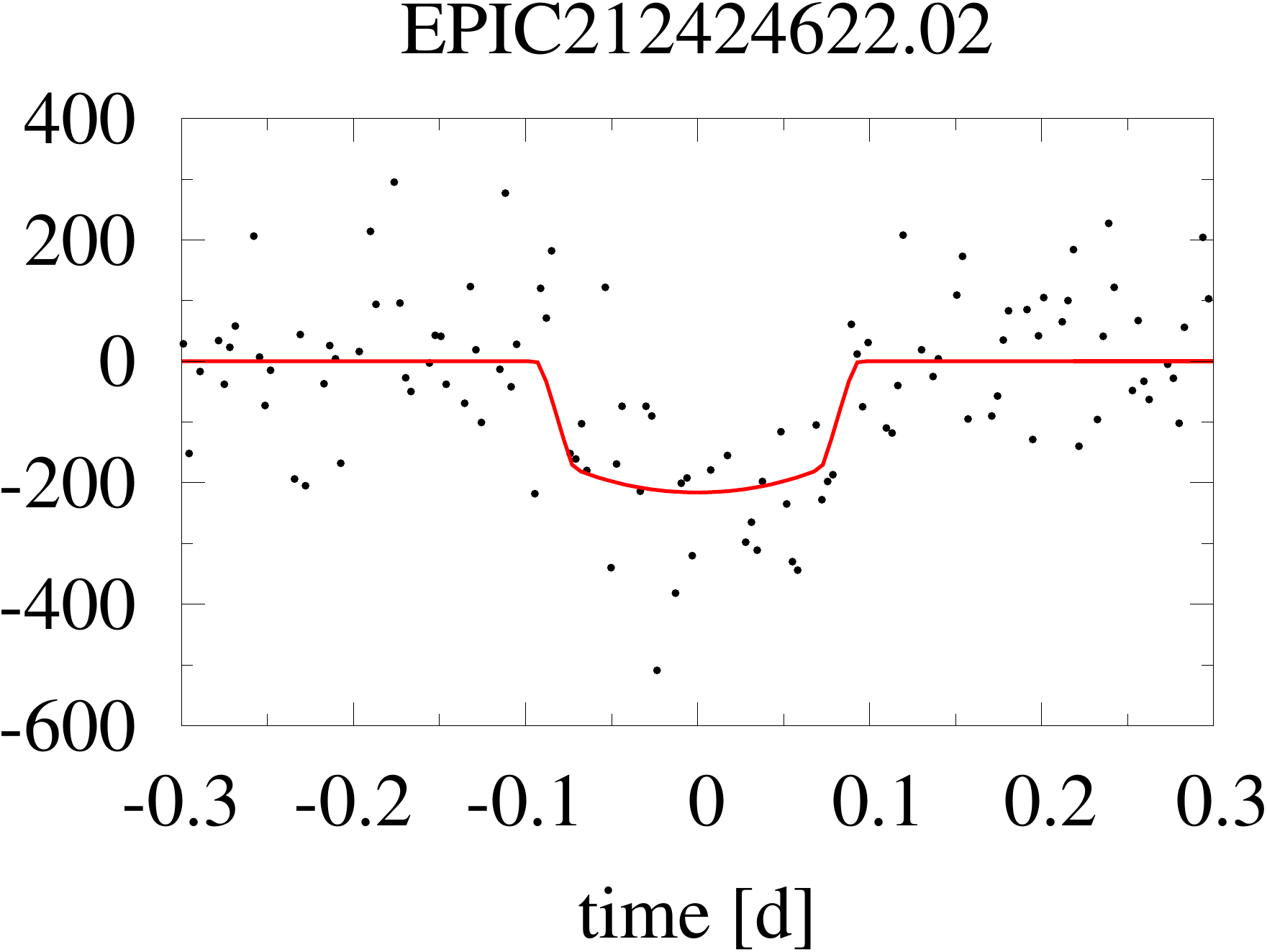}
\includegraphics[width=0.242\linewidth]{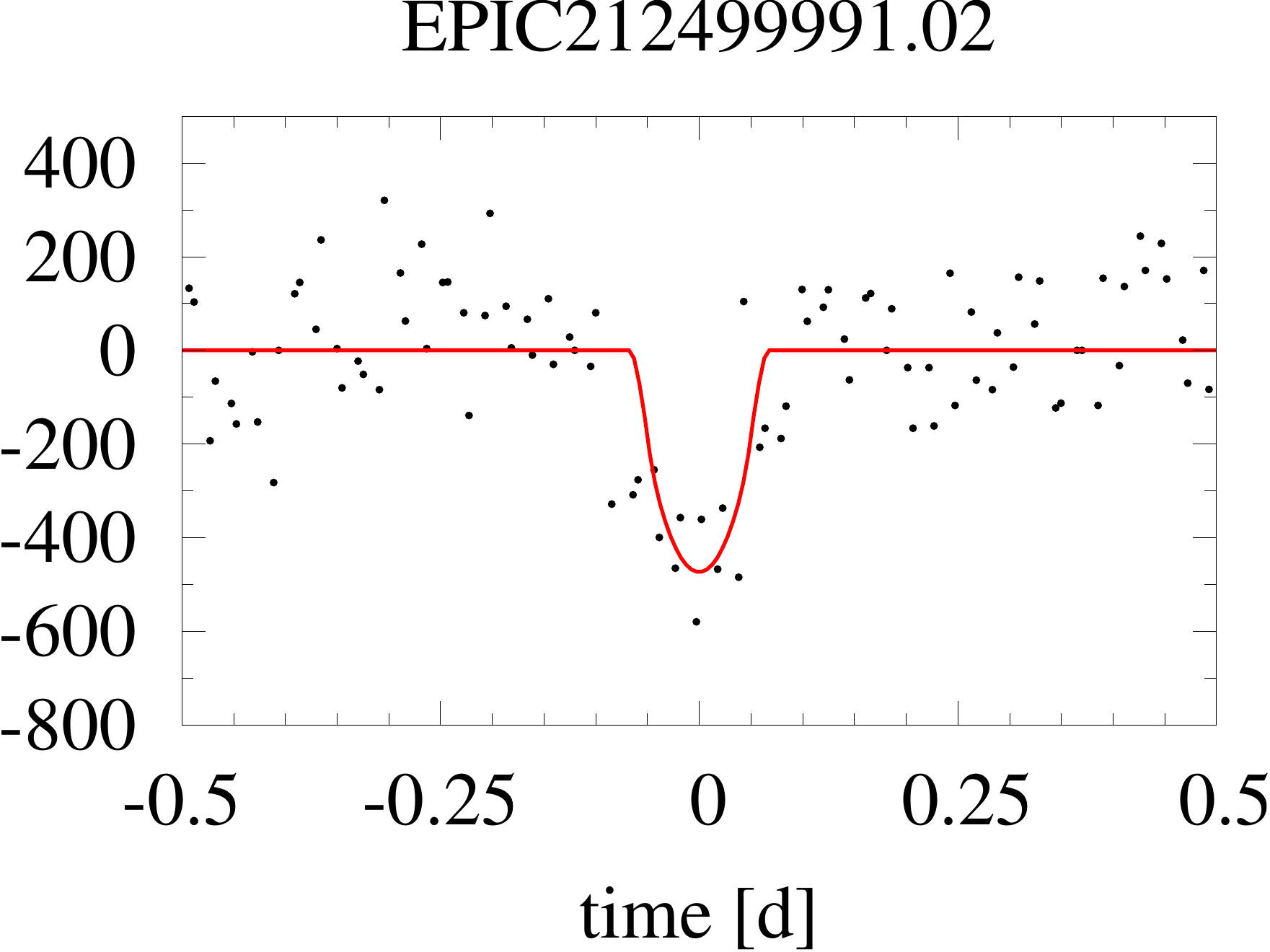}
\includegraphics[width=0.242\linewidth]{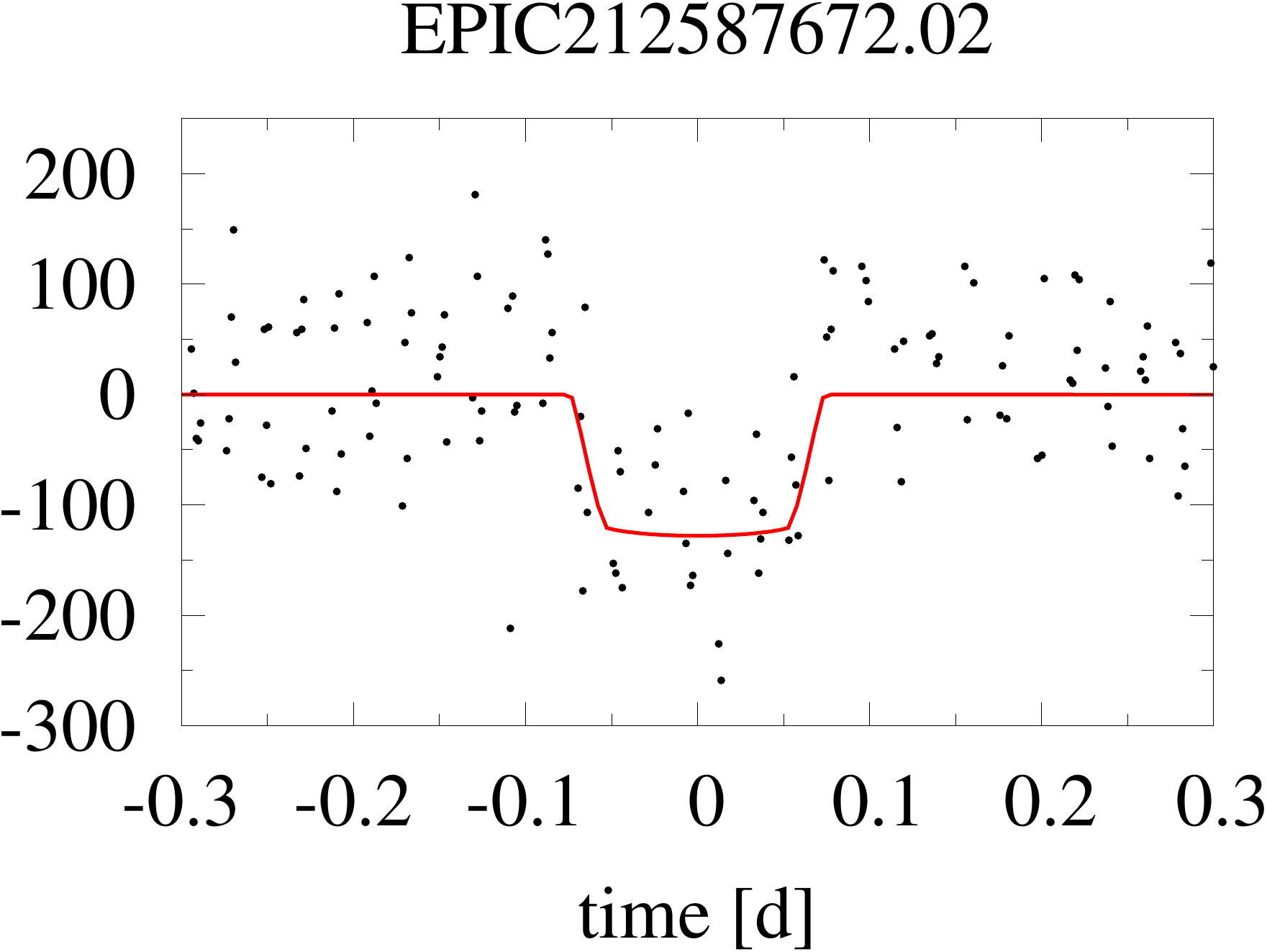}\\
\vspace{.22cm}
\includegraphics[width=0.265\linewidth]{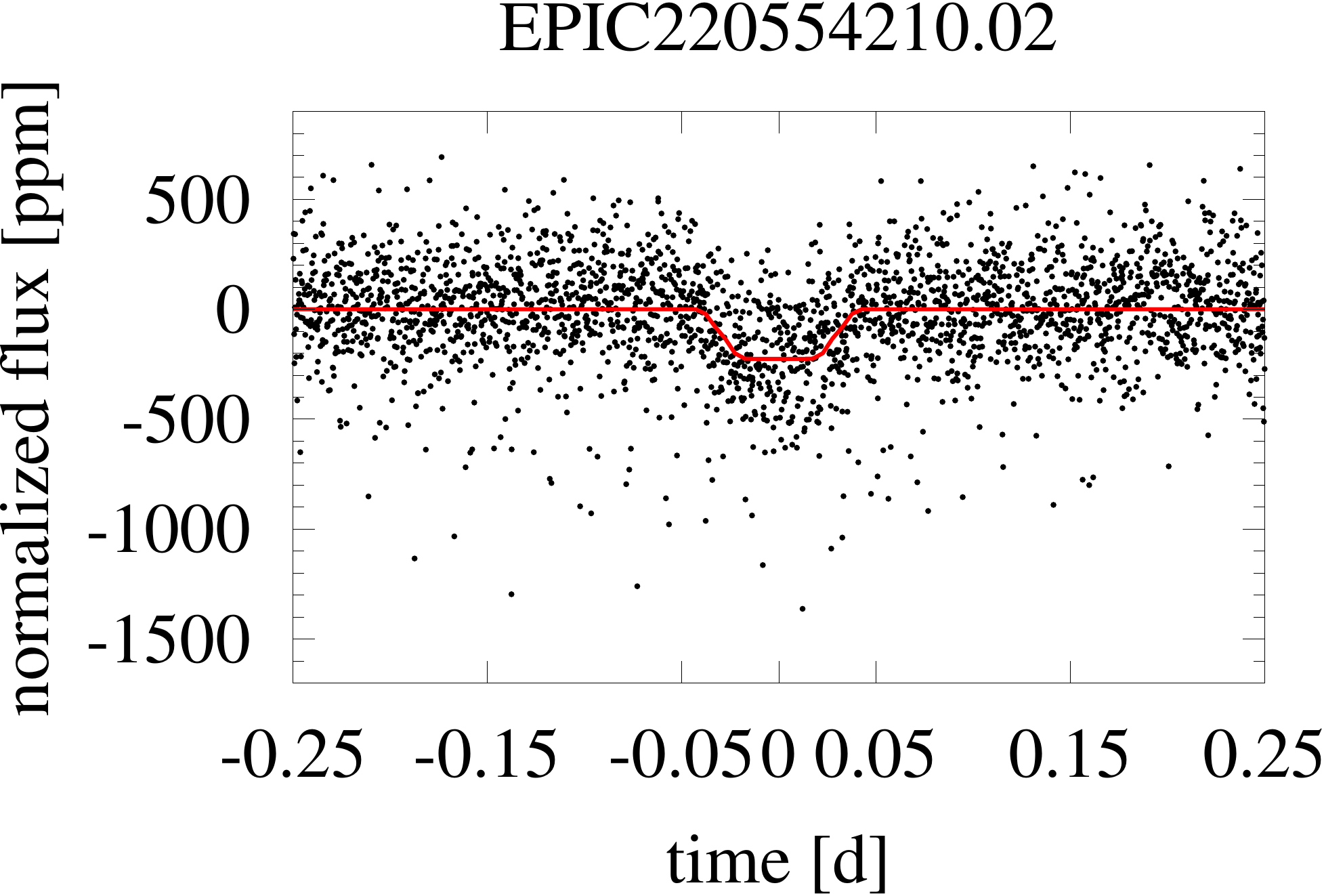}
\caption{Gallery of the phase-folded transit light curves for each of the 17 newly discovered planets.}
\label{fig:gallery}
\end{figure*}

Our vetting process includes an automated detection of transit candidates with {\tt TLS} and a visual inspection of the light curve together with the derived basic properties of the candidate.

The automated part of the vetting pipeline uses several vetting criteria, the most important of which is the signal detection efficiency of {\tt TLS} (SDE$_{\rm TLS}$). As shown by \citet{2019A&A...623A..39H}, an SDE$_{\rm TLS}$ of 9 would result in a false-positive rate $<10^{-4}$ in the limiting case of white noise. Of course, {\it K2} light curves contain various sources of time-dependent variability so that we expected (and indeed found) a higher false-positive rate. If a candidate exhibited five or fewer transits, then all transits were required to have a temporal separation of at least 0.5\,d from the beginning or the end of any gaps in the respective light curve to avoid false positives created by instrumental trends and by the detrending procedures. If a candidate exhibited three or fewer transits, then a signal-to-noise ratio (S/N) $>\,10$ was required, which we computed as ${\rm S/N}=(\delta/\sigma_{\rm o})n^{1/2}$ with $\delta$ as the mean transit depth, $\sigma_{\rm o}$ as the standard deviation of the out-of-transit points, and $n$ as the number of in-transit data points \citep{2006MNRAS.373..231P}.

Furthermore, for all objects we required that the average depth of the odd and even transits agreed within $<3\,\sigma$. We also ignored any objects with evidence of a secondary eclipse at the $>3\,\sigma$ level compared to the local noise at half an orbital phase after the candidate transit. The latter two conditions were meant to reject eclipsing binaries. In our search for transiting planets, which we suspected to be much smaller than Jupiter or Neptune (because otherwise they would have been found previously), we also rejected any candidates that show signs of phase-curve variations in the vetting sheet. This signature is much more likely to be caused by an eclipsing stellar binary than by an Earth- or super-Earth-sized object.

For any object that passed all our vetting criteria, our pipeline automatically created a vetting sheet that we used for our visual inspection. Figure~\ref{fig:vetting} presents an example, and the panels are explained in the figure caption. The Lomb-Scargle periodograms shown in the bottom panels were computed with the implementation of {\tt astropy,} which makes use of the floating mean periodogram of \citet{2009A&A...496..577Z}.


\subsection{False-positive probabilities}

We used the {\tt vespa} software \citep{2012ApJ...761....6M,2015ascl.soft03011M} to evaluate the false-positive probability (FPP) for each of our candidates. Details of our usage of {\tt vespa} are given in \PaperI. In brief, we used the star's astrophysical characterization, that is, its effective temperature, surface gravity, and metallicity, from the K2 Ecliptic Plane Input Catalog of \citet{2016ApJS..224....2H} and the astronomical information from the 2MASS broadband photometry \citep[$J$, $H$, $K$;][]{2003yCat.2246....0C}. The stellar radii for EPIC\,201497682, EPIC\,201841433, EPIC\,212297394, and EPIC\,212499991 listed by \citet{2016ApJS..224....2H} have uncertainties of up to several solar radii. For these objects we used the stellar radii published in the Gaia data release 2 (DR2) \citep{2016A&A...595A...1G,2018A&A...616A...1G} as estimated by \citet{2018A&A...616A...8A} in order to compute the planetary radii from the respective stellar radius and the planet-to-star radius ratio measured with our Markov chain Monte Carlo (MCMC) fitting (see Sect.~\ref{sec:MCMC}).

One critical criterion for the computation of the FPP with {\tt vespa} was the maximum aperture radius ($\rho$) interior to which the signal must be produced (the maxrad parameter in {\tt vespa}). We retrieved and inspected optical images from the Panoramic Survey Telescope and Rapid Response System \citep[Pan-STARRS;][]{2016arXiv161205560C} through the {\tt Aladin} software to evaluate the possibility of contamination by uncorrelated objects and to determine $\rho$. Pan-STARRS is complete to visual magnitudes of about 22 with a resolution down to 200\,mas per pixel and therefore both sufficiently deep and sufficiently resolved for us to constrain the presence of contaminants. Blurring effects from astronomical seeing and the Pan-STARRS telescope point spread function limited our resolution to a minimum of $\rho~{\sim}~1\,\arcsec$ in those cases where no resolved sources of contamination were visible. One notable exception was EPIC\,201754305 (K2-16), for which we referred to adaptive optic imaging from \citet{2016ApJ...827...78S} and restricted contamination to within $\rho=0.5\,\arcsec$.

The {\it K2} photometric apertures of our targets were typically smaller than about six pixels or $\leq\,24\,\arcsec$ given that the spatial resolution of the {\it K2} CCD is $4\,\arcsec$ per pixel. For all our candidates, we compared the corresponding {\it K2} Full Frame Images\footnote{Available at \href{https://archive.stsci.edu/k2/ffi_display.php}{https://archive.stsci.edu/k2/ffi\_display.php}} of the respective {\it K2} campaign and CCD channel as well as the {\it K2} CCD images of their aperture masks used by {\tt EVEREST} with Pan-STARRS images to search for stellar contaminants within the respective apertures. In Fig.~\ref{fig:PANSTARRS_contaminants} we show the corresponding images that we inspected for EPIC\,206215704 (top left), EPIC\,206317286 (top right), EPIC\,212424622 (bottom left)\footnote{A structure nicknamed ``Fuzzy'' that was likely caused by a piece of fabric on the CCD when collecting laboratory-based data for the flat-field correction (Doug Caldwell \& Geert Barentsen, priv. comm.) is labeled and marked with a curved light blue arrow.}, and EPIC\,212499991 (bottom right) as examples. Initial examination of the Pan-STARRS images suggested the presence of nearby possible stellar contaminants to these objects, but the comparison with the correctly oriented {\it K2} postage stamps\footnote{Available on the Data Validation Summary through the python implementation of {\tt EVEREST} via\\
{\tt
>import everest\\
>everest.DVS(206215704) 
}} showed that they are outside of the respective aperture masks of EPIC\,206215704, EPIC\,206317286, and EPIC\,212424622. In these cases we concluded that no contaminant is optically resolved from the three targets and that the measured flux must have been produced within the respective point spread functions ($\rho=1\,\arcsec$) of the targets within the corresponding Pan-STARRS images. In the fourth case of EPIC\,212499991, we found indeed that the candidate contaminant is within the {\it K2} aperture mask used by {\tt EVEREST}. The apparent separation between the target and this nearby source is $\rho~=~11.1\,\arcsec$ , and {\tt vespa} predicted an FPP of $2\times10^{-1}$.

\begin{figure*}[h!]
\centering
\vspace{-.24111cm}
\includegraphics[width=.48\linewidth]{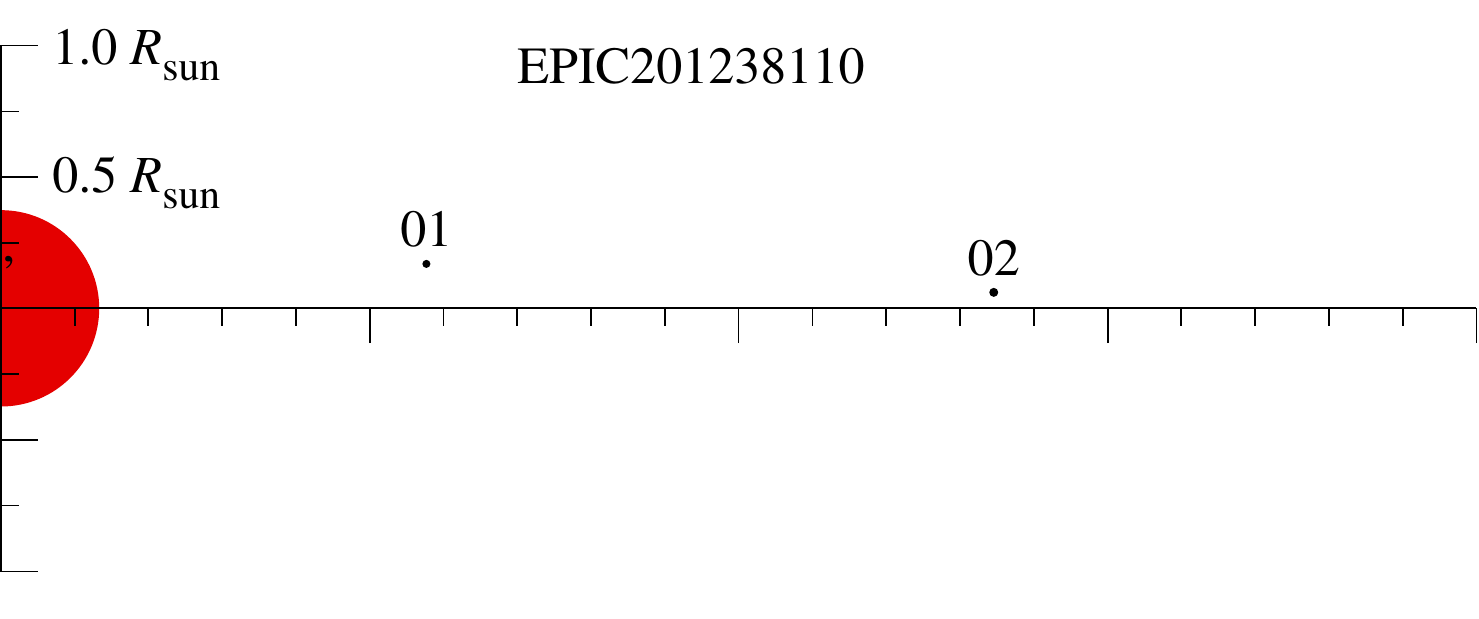}
\hspace{0.3333cm}
\includegraphics[width=.48\linewidth]{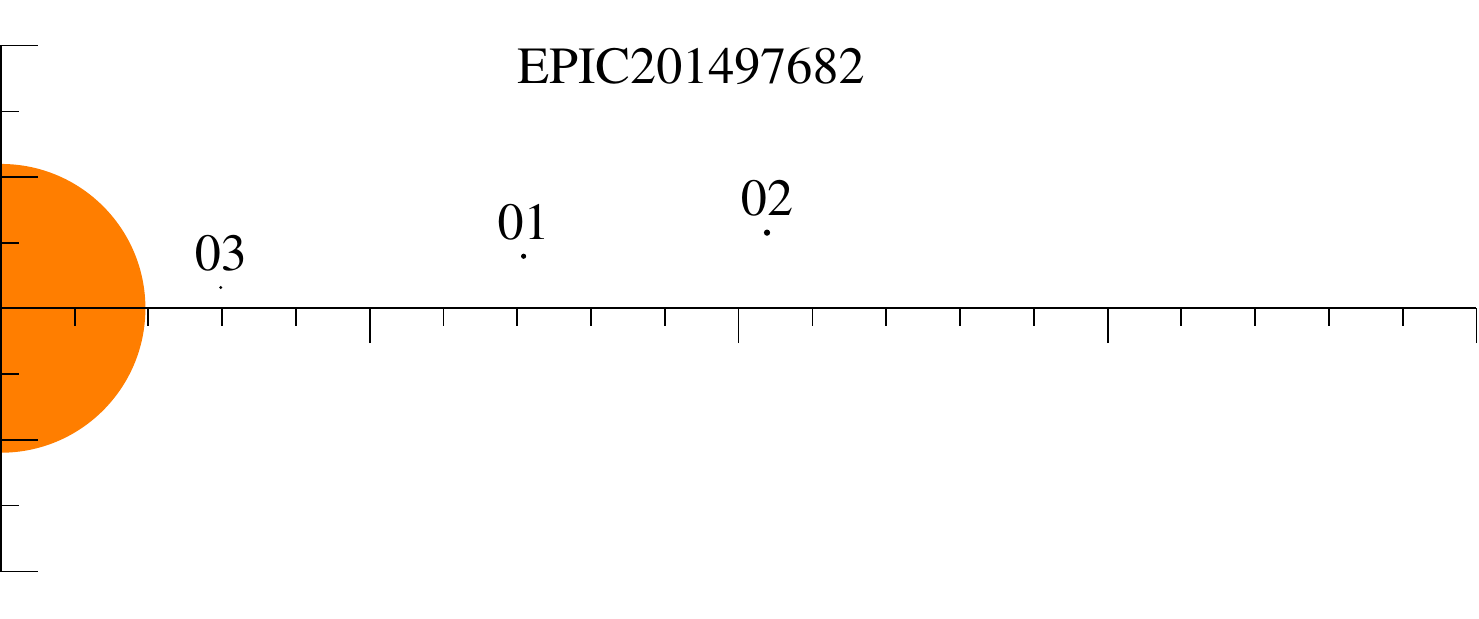}\\
\vspace{-.26111cm}
\includegraphics[width=.48\linewidth]{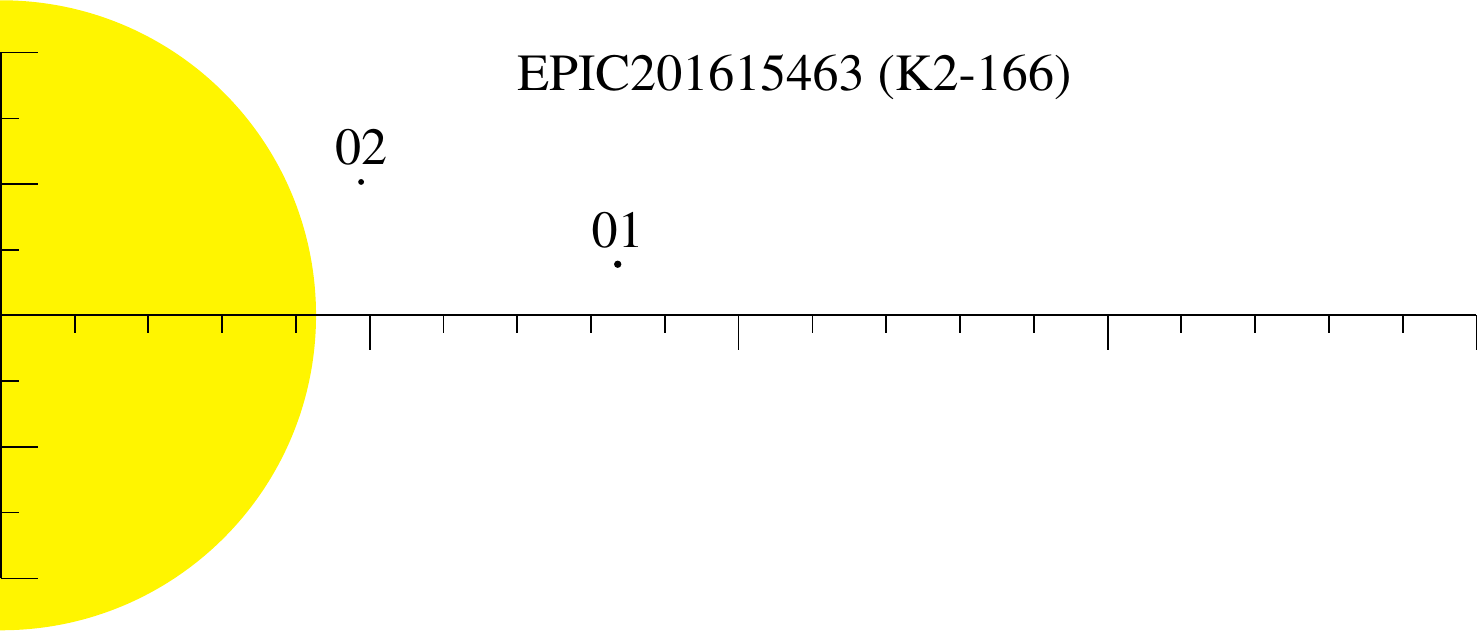}
\hspace{0.3333cm}
\includegraphics[width=.48\linewidth]{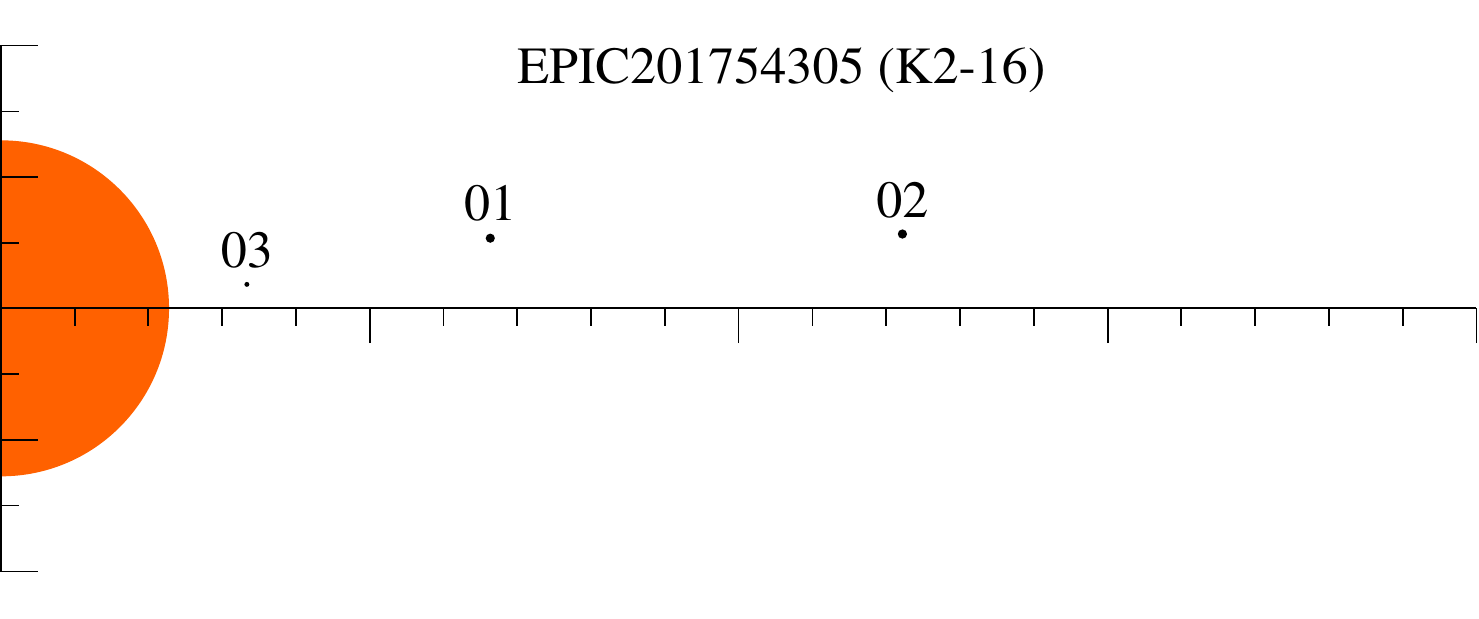}\\
\vspace{-.26111cm}
\includegraphics[width=.48\linewidth]{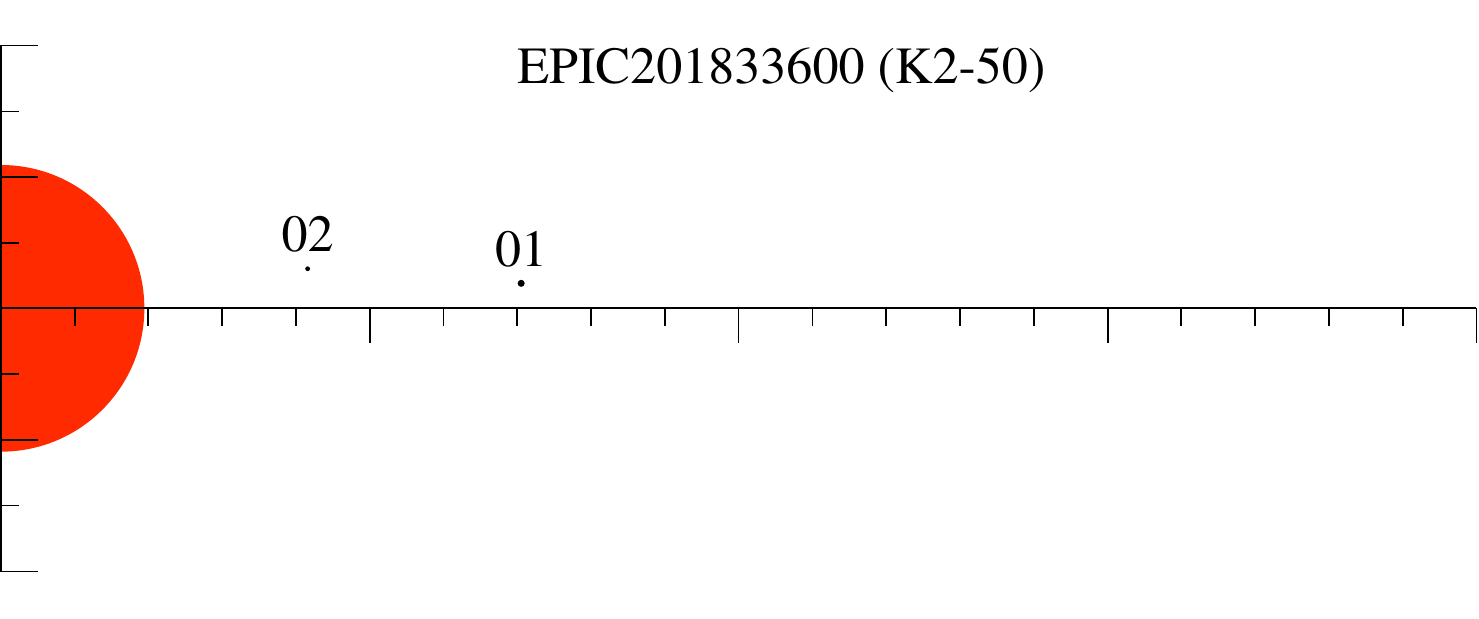}
\hspace{0.3333cm}
\includegraphics[width=.48\linewidth]{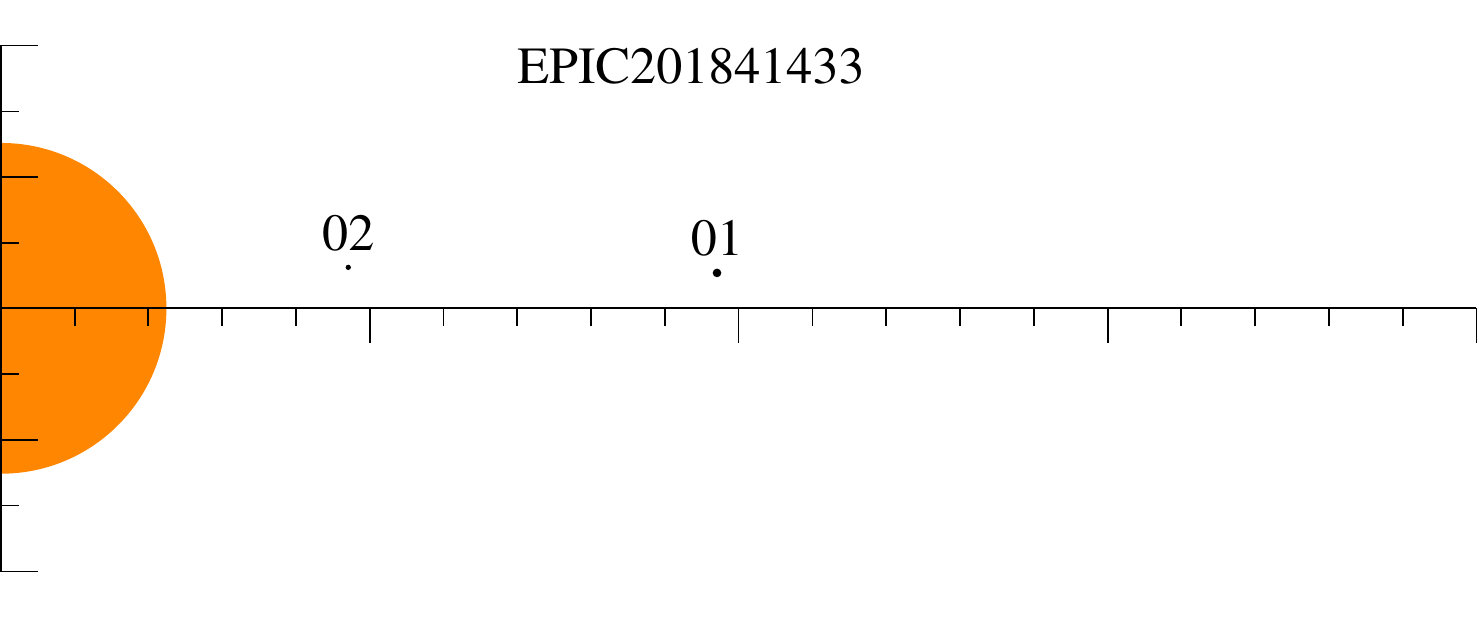}\\
\vspace{-.26111cm}
\includegraphics[width=.48\linewidth]{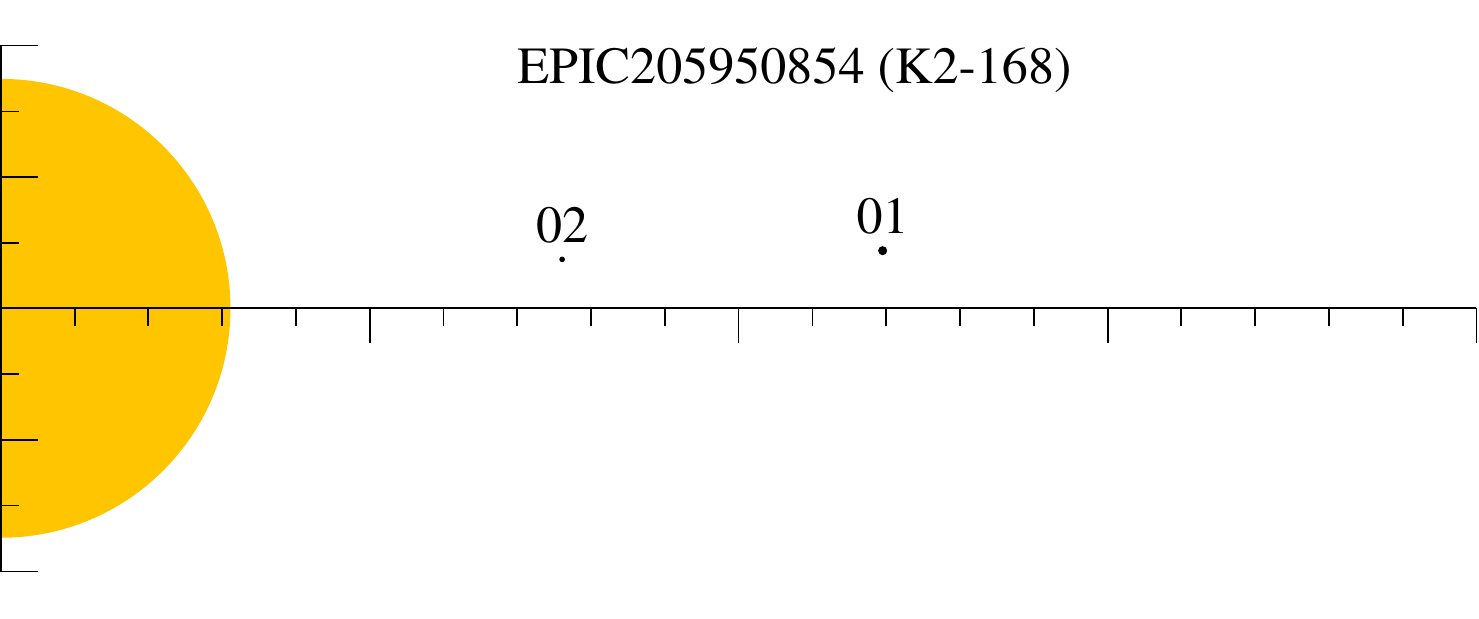}
\hspace{0.3333cm}
\includegraphics[width=.48\linewidth]{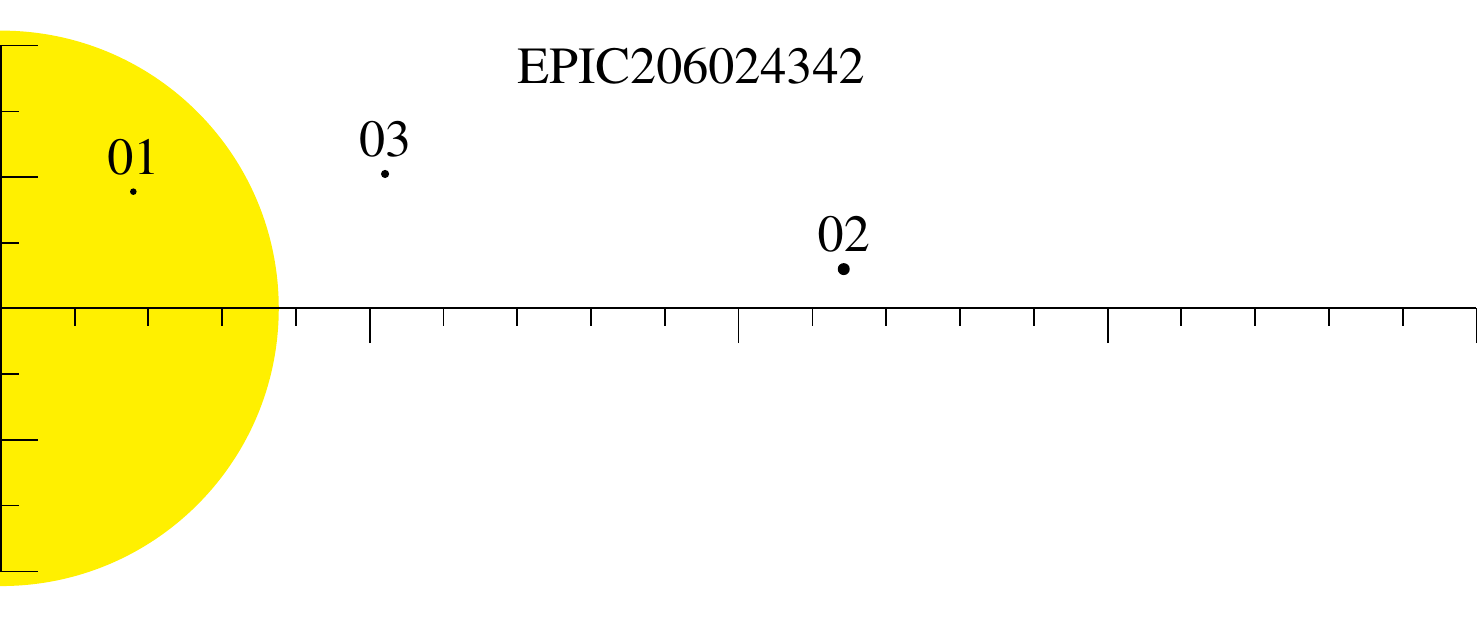}\\
\vspace{-.26111cm}
\includegraphics[width=.48\linewidth]{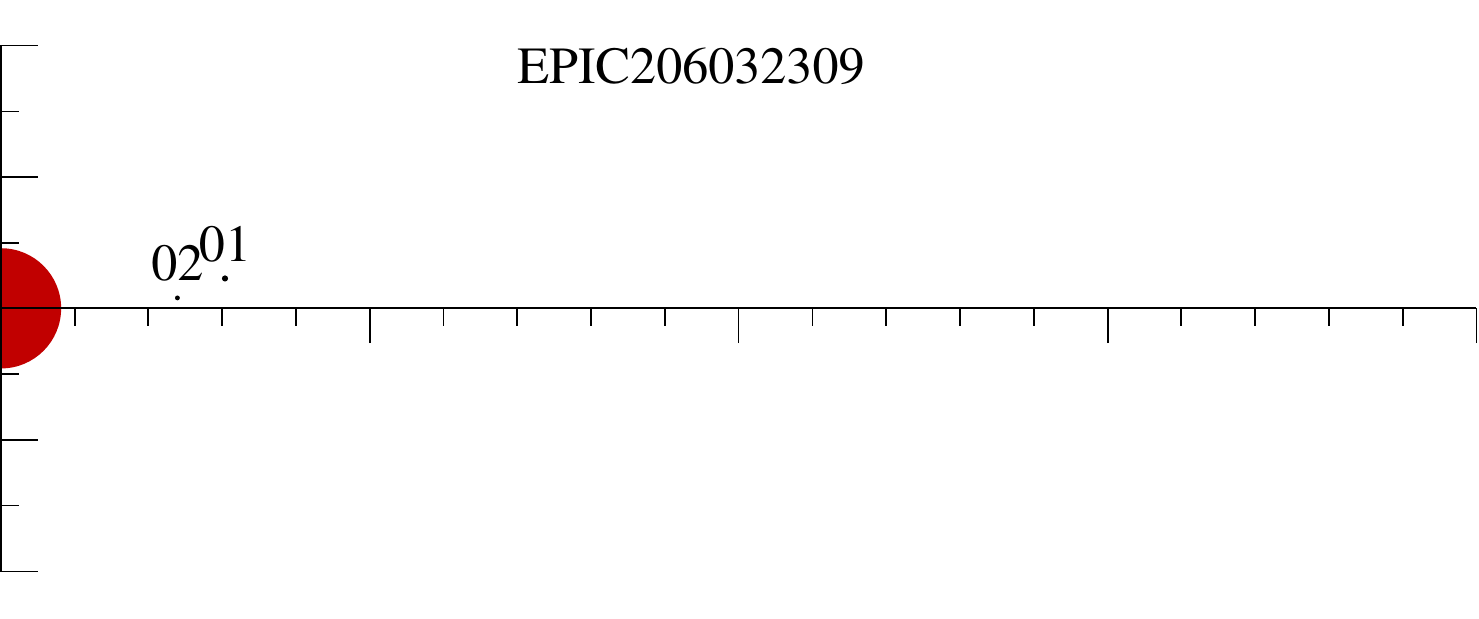}
\hspace{0.3333cm}
\includegraphics[width=.48\linewidth]{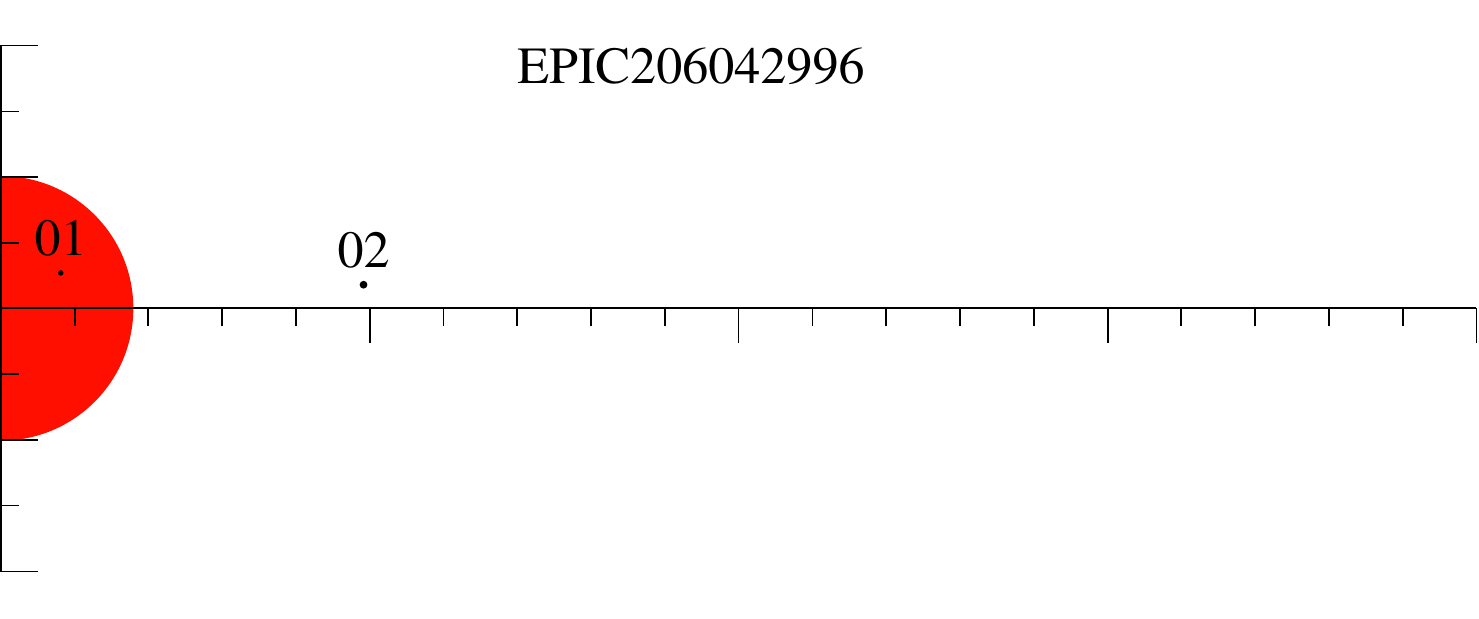}\\
\vspace{-.26111cm}
\includegraphics[width=.48\linewidth]{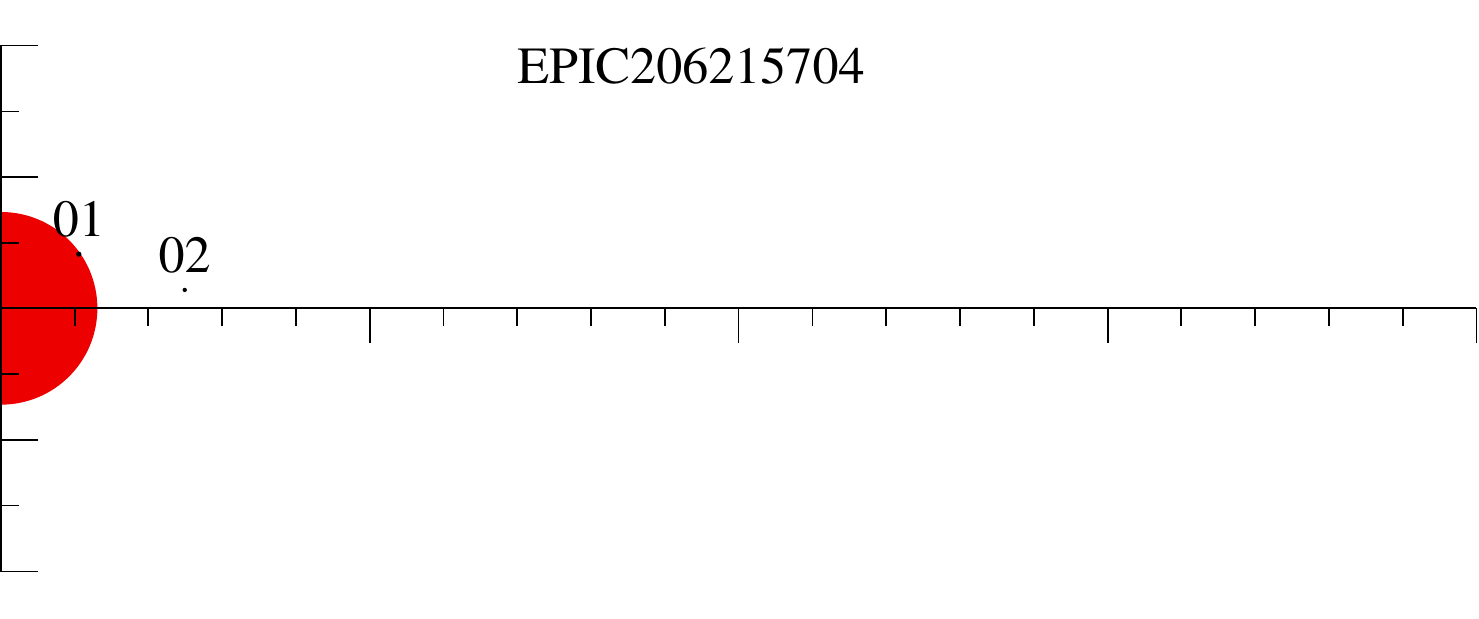}
\hspace{0.3333cm}
\includegraphics[width=.48\linewidth]{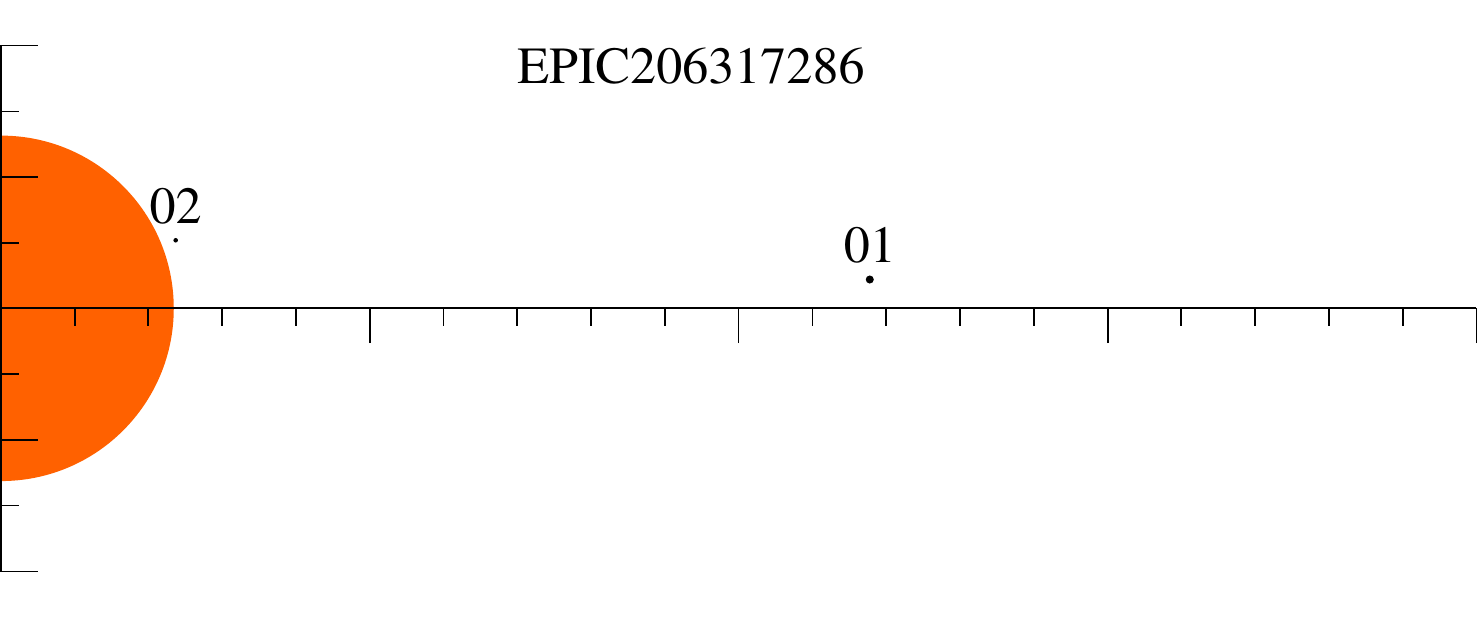}\\
\vspace{-.26111cm}
\includegraphics[width=.48\linewidth]{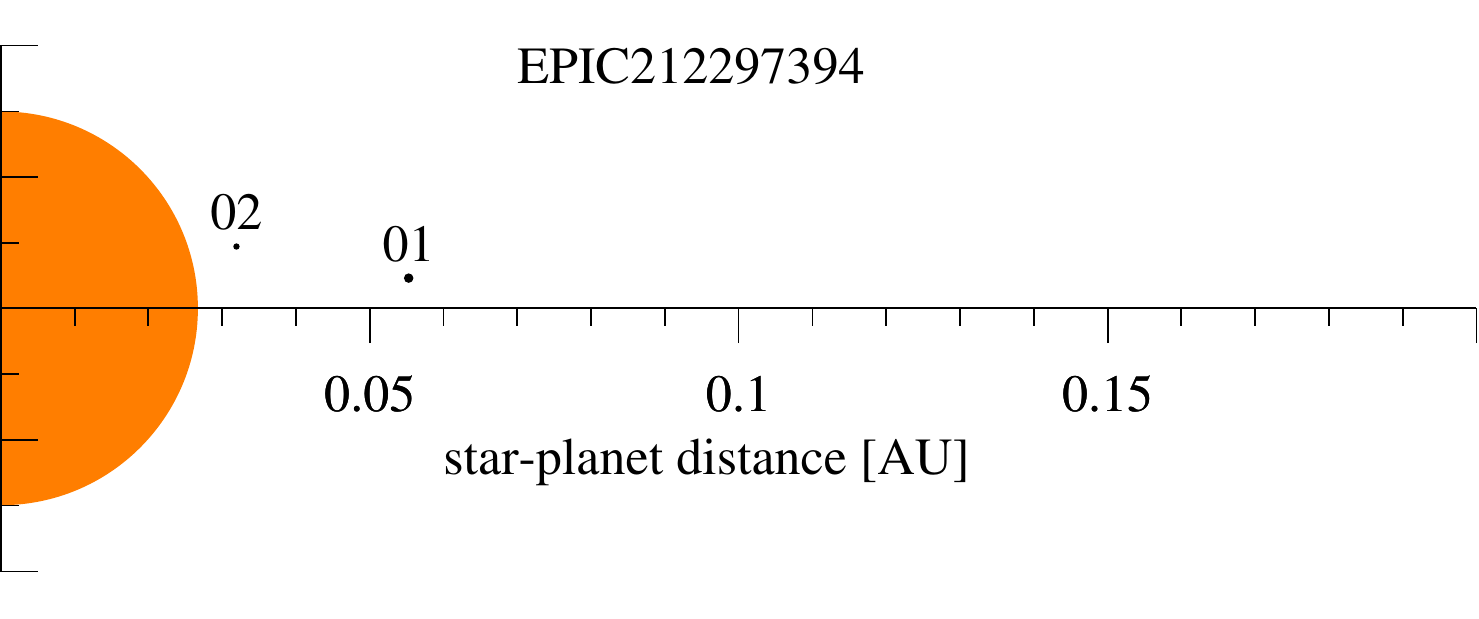}
\hspace{0.3333cm}
\includegraphics[width=.48\linewidth]{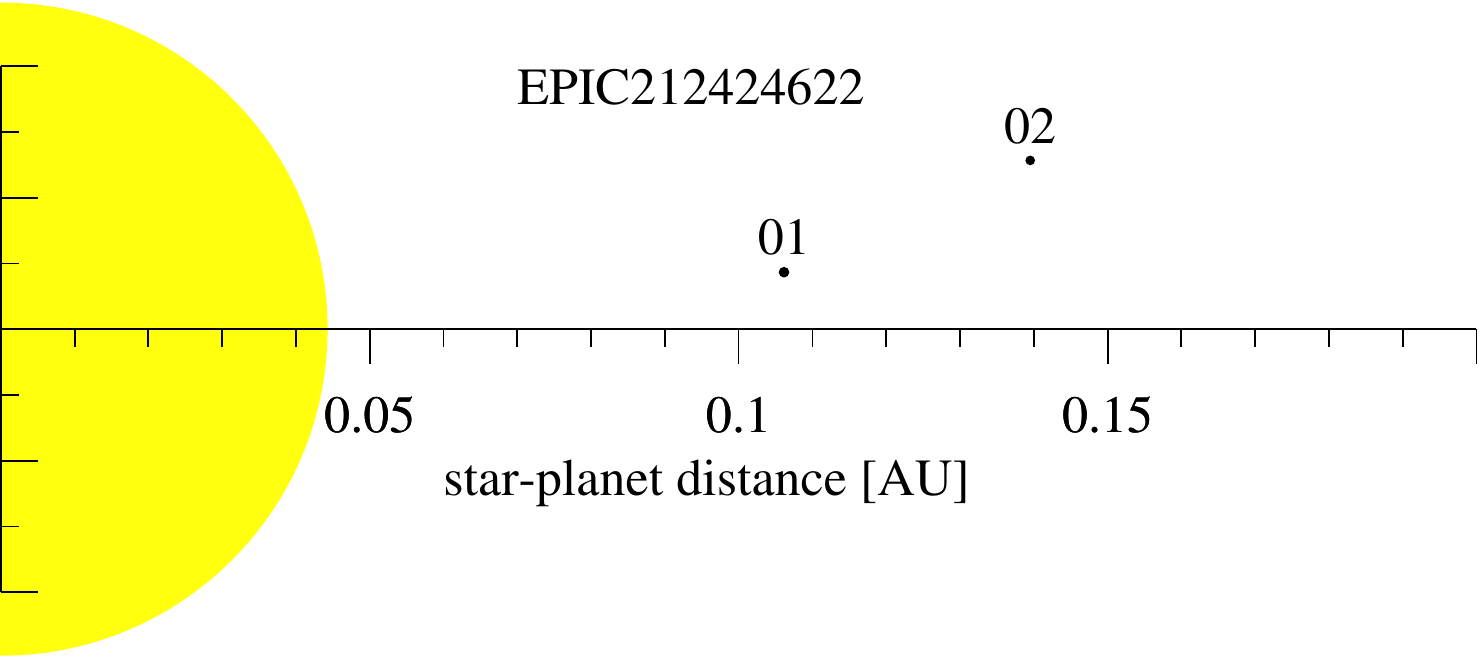}
\vspace{-.21111cm}
\caption{{\it (continued on next page)}}
\end{figure*}

\setcounter{figure}{4}

\begin{figure*}[h!]
\centering
\includegraphics[width=.48\linewidth]{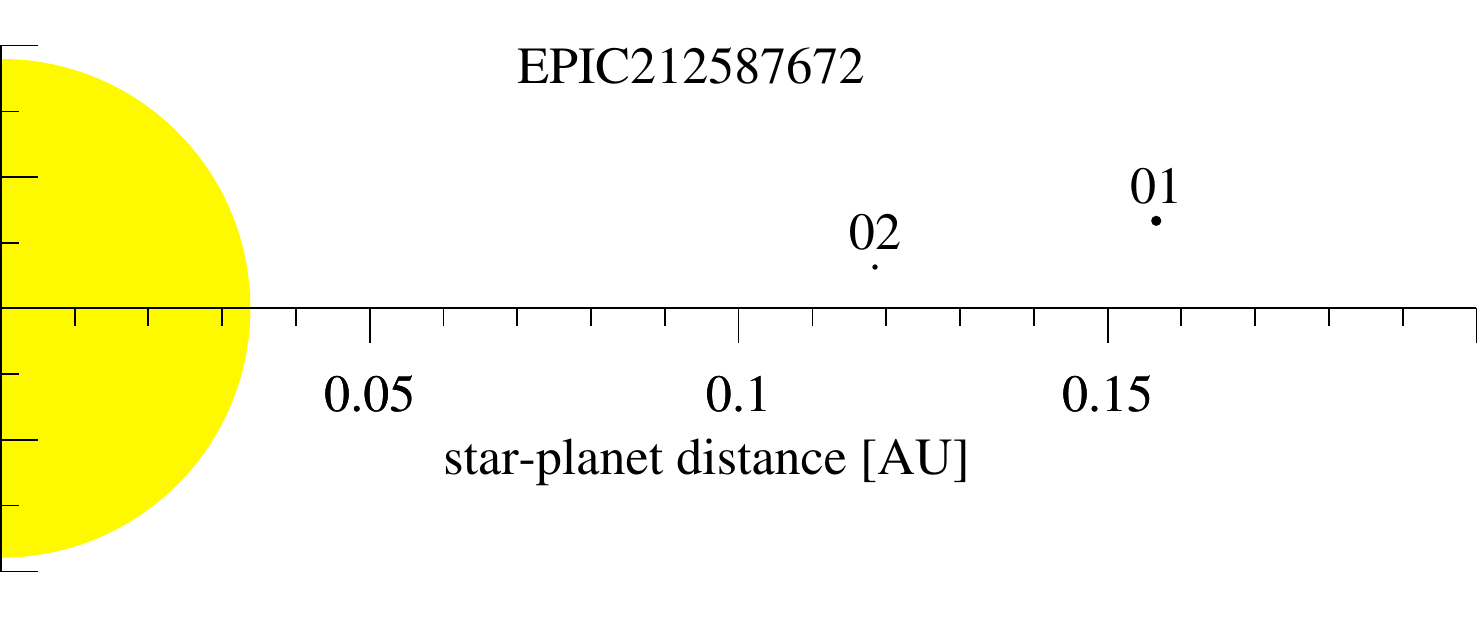}
\hspace{0.3333cm}
\includegraphics[width=.48\linewidth]{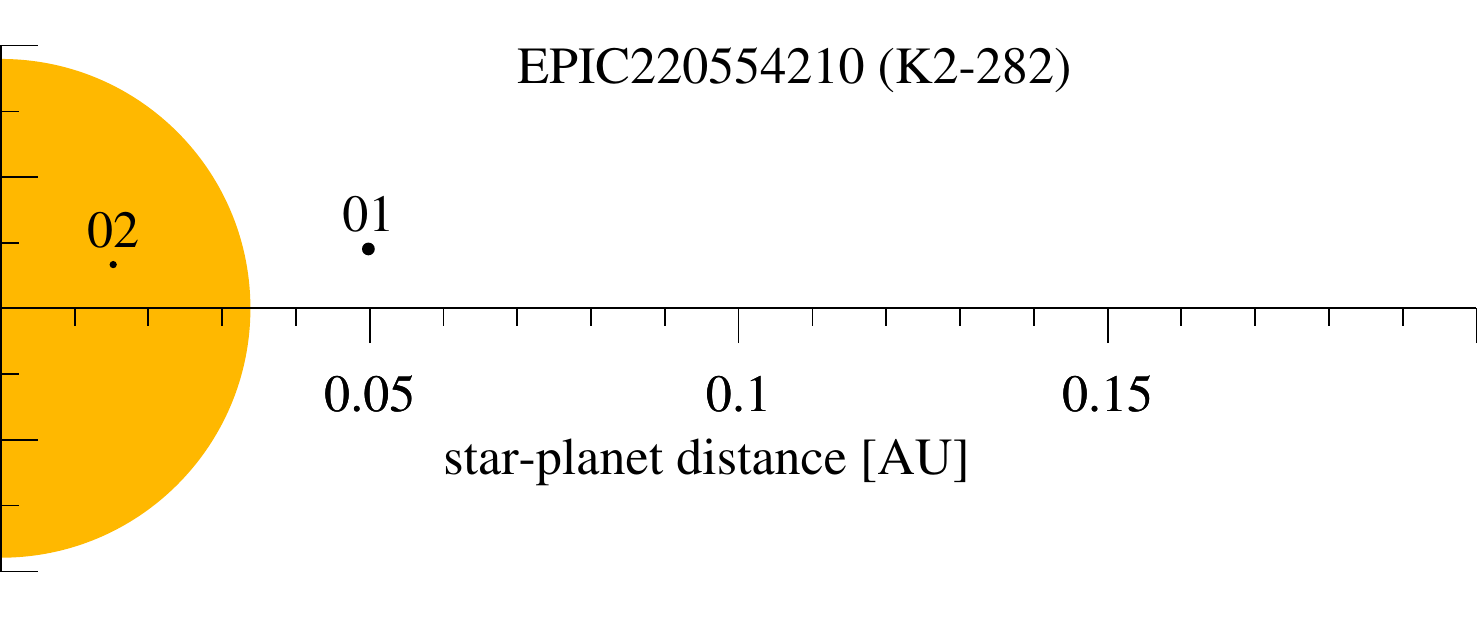}\\
\vspace{-.2222cm}
\includegraphics[width=.51\linewidth]{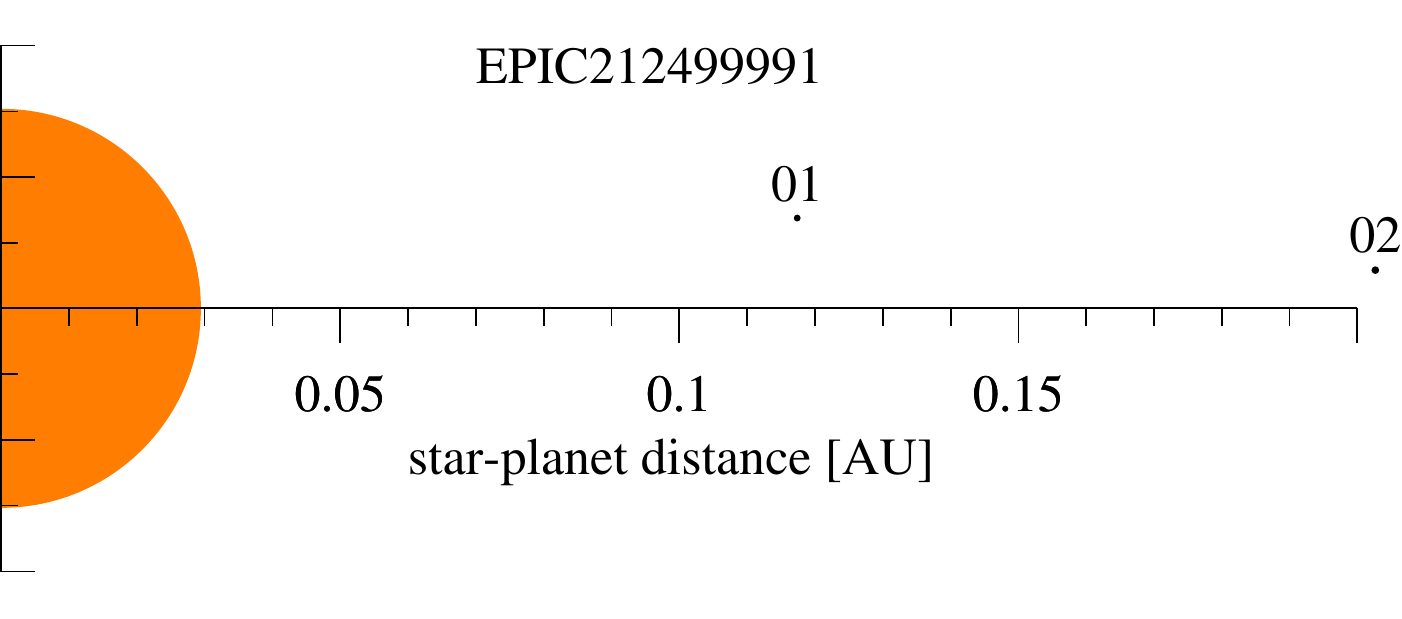}
\caption{Orbital architectures of the 17 systems with new planets. The colors of the stars indicate their effective temperatures, ranging from 3500\,K (red) to 6100\,K (yellow). All planets known in each system are shown, where the new planets are designated with the highest number (02 or 03) in each system. The positions of the planets are chosen based on the results of our MCMC fitting of the entire {\it K2} light curves, with the $x$ coordinate corresponding to the star-planet distance (in AU) and the $y$ coordinate relating to the transit impact parameter. The ordinate runs from zero to the radius of the Sun. All stellar and planetary radii are to scale. Star-planet distances are also to scale, but radii are not to scale with distances.}
\label{fig:architectures}
\end{figure*}

We here present only objects for which we derived an FPP lower than 1\,\%. We designate these objects as ``statistically validated'' exoplanets. As a consequence, with an ${\rm FPP}=2\times10^{-1}$ EPIC\,212499991 cannot be validated with {\tt vespa} and our constraints on the possibility of nearby contaminants. We can, however, make use of the ``multiplicity boost'' factor ($X_2$) that is implied by the presence of two planet candidates within the same aperture. These candidates have much lower FPPs than single candidates \citep{2012ApJ...750..112L}. We computed the multiplicity-corrected FPP for a second planet (FPP$_2$) from the probability of planethood in a two-planet system ($P_2$) as ${\rm FPP}_2 = 1-P_2$, where

\begin{equation}\label{eq:fpp2}
P_2 \approx \frac{X_2 P_1}{X_2 P_1 + (1-P_1),}
\end{equation}

\noindent
with $P_1 = 1-{\rm FPP}$ and $X_2=25$, as estimated by both \citet{2012ApJ...750..112L} and \citet{2016ApJ...827...78S}. For any system in which we detected a third transiting candidate rather than a second, the multiplicity boost is even stronger and the resulting FPP for that specific object is lower than the values that we present. In the case of EPIC\,212499991, the multiplicity boost derived with Eq.~\eqref{eq:fpp2} pushed the transit candidate below the validation threshold with an ${\rm FPP}_2=9.9~\times~10^{-3}$. All other planets in our sample were found to have substantially lower FPP$_2$ values.

\subsection{System characterization with MCMC fitting}
\label{sec:MCMC}

As in \PaperI, we used the MCMC sampler {\tt emcee} \citep{2013PASP..125..306F} to constrain the planetary parameters. In brief, we provided {\tt emcee} with the times of the mid-points of the first transit ($T_1$) and with the orbital period ($P$) as measured with {\tt TLS} for each planet in a given system. For each planet, $P$, $T_1$, the planet-to-star radius ratio ($r_{\rm p}=R_{\rm p}/R_{\rm s}$) and the transit impact parameter ($b$) were free model parameters, whereas the stellar density ($\rho_{\rm s}$) and the two limb-darkening coefficients \citep[according to the quadratic limb-darkening law of][]{2013MNRAS.435.2152K} were global parameters for all transits in a given light curve. Each of the 100 MCMC walkers applied to a light curve executed $200,000$ steps, the first half of which we neglected to only preserve burned-in MCMC chains.

We measured the time of mid-transit of the first transit after the center of the respective {\it K2} light curve for a given object, $T_0~=~{\rm BKJD}-2065$\,d, where the Barycentric Kepler Julian Day is BKJD = BJD - 2,454,833.0\,d. This value minimizes the error in $T_0$ and it minimizes the correlation between $T_0$ and $P$.

To deduce the errors ($\sigma_{R_{\rm p}}$) of the planetary radii $R_{\rm p}~=~r_{\rm p}R_{\rm s}$, we propagated the respective error of the stellar radius ($\sigma_{R_{\rm s}}$) reported in the literature and the error in the planet-to-star radius ratio obtained with our MCMC fitting ($\sigma_{r_{\rm p}}$) through a first-order Taylor series expansion,

\begin{equation}
\sigma_{R_{\rm p}}~=~\sqrt{( {\partial}R_{\rm p}/{\partial}r_{\rm p})^2 \sigma_{r_{\rm p}}^2 + ( {\partial}R_{\rm p}/{\partial}R_{\rm s})^2 \sigma_{R_{\rm s}}^2}=\sqrt{R_{\rm s}^2\sigma_{r_{\rm p}}^2 + r_{\rm p}^2\sigma_{R_{\rm s}}^2} \ .
\end{equation}

\section{Results}
\label{sec:results}

\begin{sidewaystable*}
\caption{Characterization and validation of the 17 new planets from our {\tt TLS} search, MCMC fitting, and {\tt vespa} false-positive vetting pipeline.}
\def\arraystretch{2}
\label{tab:MCMC}
\small
\centering
\begin{tabular}{c c c c c c c c c c c c c c}
\hline\hline
EPIC & {\it K2} name & $M_{\rm s}\,[M_\odot]$ & $T_0$\tablefootmark{(a)} & $P$\,[d] & $R_{\rm p}/R_{\rm s}$ & $R_{\rm p}\,[R_\oplus]$ & $b$ & \#tr\tablefootmark{(b)} & SNR & SDE$_{\rm TLS}$ & $\rho\,[$\arcsec$]$ & FPP & FPP$_2$ \\
\hline
201238110.02\tablefootmark{(d)} &          & $0.410_{-0.048}^{+0.112}$ & $-45.3428_{-0.0023}^{+0.0023}$ & $28.1656_{-0.0028}^{+0.0027}$ & $0.0458_{-0.0022}^{+0.0021}$ & $1.87_{-0.45}^{+0.20}$ & $0.24_{-0.17}^{+0.24}$ & 3 & 12.1 & 27.5 & $1.0$ & $5.3~\times~10^{-4}$ & $2.1~\times~10^{-5}$\\
201497682.03\tablefootmark{(d)} &          & $0.780_{-0.170}^{+0.085}$ & $-47.9737_{-0.0026}^{+0.0028}$ & $2.13174_{-0.00022}^{+0.00022}$ & $0.0115_{-0.0008}^{+0.0008}$ & $0.692_{-0.048}^{+0.059}$ & $0.21_{-0.15}^{+0.23}$ & 36 & 9.5 & 21.2 & 22.3 & $5.0~\times~10^{-2}$ & $2.1~\times~10^{-3}$ \\
201615463.02\tablefootmark{(i)} &  K2-166  & $1.073_{-0.034}^{+0.034}$ & $723.2227_{-0.0053}^{+0.0051}$ & $3.80464_{-0.00105}^{+0.00091}$ & $0.0093_{-0.0007}^{+0.0007}$ & $1.22_{-0.17}^{+0.21}$ & $0.61_{-0.15}^{+0.14}$ & 15 & 8.0 & 17.7 & 24.1 & $1.7~\times~10^{-2}$ & $6.8~\times~10^{-4}$ \\
201754305.03\tablefootmark{(cde)} &  K2-16   & $0.670_{-0.040}^{+0.030}$ & $-47.5037_{-0.0029}^{+0.0029}$ & $2.71578_{-0.00031}^{+0.00031}$ & $0.015_{-0.001}^{+0.001}$ & $1.03_{-0.086}^{+0.085}$ & $0.21_{-0.14}^{+0.21}$ & 26 & 9.6 & 22.9 & 0.5 & $3.2~\times~10^{-5}$ & $1.3~\times~10^{-6}$ \\
201833600.02\tablefootmark{(de)} &  K2-50   & $0.611_{-0.056}^{+0.056}$ & $-45.3448_{-0.0025}^{+0.0025}$ & $3.96151_{-0.00051}^{+0.00046}$ & $0.0168_{-0.0012}^{+0.0012}$ & $1.00_{-0.11}^{+0.12}$ & $0.40_{-0.24}^{+0.20}$ & 18 & 8.0 & 12.7 & $10.0$ & $8.3~\times~10^{-3}$ & $3.3~\times~10^{-4}$ \\
201841433.02\tablefootmark{(d)} &          & $0.802_{-0.163}^{+0.081}$ & $-44.6700_{-0.0033}^{+0.0037}$ & $4.16959_{-0.00053}^{+0.00051}$ & $0.0160_{-0.0016}^{+0.0017}$ & $1.10_{-0.12}^{+0.14}$ & $0.36_{-0.24}^{+0.25}$ & 18 & 7.3 & 19.1 & 1.0 & $3.8~\times~10^{-3}$ & $1.5~\times~10^{-4}$ \\
205950854.02\tablefootmark{(di)} &  K2-168  & $0.907_{-0.029}^{+0.032}$ & $115.9808_{-0.0061}^{+0.0062}$ & $8.0468_{-0.0025}^{+0.0024}$ & $0.0124_{-0.0009}^{+0.0010}$ & $1.18_{-0.10}^{+0.13}$ & $0.31_{-0.21}^{+0.25}$ & 9 & 10.0 & 21.3 & 1.0 & $2.9~\times~10^{-3}$ & $1.2~\times~10^{-4}$ \\
206024342.03\tablefootmark{(de)} &          & $0.928_{-0.098}^{+0.082}$ & $1.1324_{-0.0027}^{+0.0026}$ & $4.50756_{-0.00060}^{+0.00062}$ & $0.0147_{-0.0010}^{+0.0012}$ & $1.70_{-0.31}^{+0.59}$ & $0.70_{-0.10}^{+0.09}$ & 13 & 11.3 & 25.1 & 27.0 & $2.4~\times~10^{-2}$ & $9.7~\times~10^{-4}$ \\
206032309.02\tablefootmark{(d)} &          & $0.221_{-0.062}^{+0.035}$ & $1.1320_{-0.0017}^{+0.0017}$ & $2.87814_{-0.00026}^{+0.00023}$ & $0.0405_{-0.0024}^{+0.0024}$ & $1.01_{-0.13}^{+0.24}$ & $0.26_{-0.18}^{+0.26}$ & 24 & 11.9 & 20.7 & 1.0 & $8.1~\times~10^{-3}$ & $3.3~\times~10^{-4}$ \\
206042996.02\tablefootmark{(d)} &          & $0.565_{-0.053}^{+0.053}$ & $1.1227_{-0.0029}^{+0.0029}$ & $5.29711_{-0.00070}^{+0.00074}$ & $0.0294_{-0.0019}^{+0.0018}$ & $1.62_{-0.19}^{+0.17}$ & $0.26_{-0.18}^{+0.23}$ & 12 & 9.7 & 19.7 & 1.0 & $8.4~\times~10^{-4}$ & $3.4~\times~10^{-5}$ \\
206215704.02\tablefootmark{(d)} &          & $0.407_{-0.079}^{+0.095}$ & $112.0005_{-0.0004}^{+0.0009}$ & $2.25372_{-0.00047}^{+0.00047}$ & $0.0225_{-0.0021}^{+0.0022}$ & $0.90_{-0.20}^{+0.19}$ & $0.28_{-0.20}^{+0.28}$ & 29 & 12.4 & 22.5 & 1.0 & $1.1~\times~10^{-1}$ & $5.0~\times~10^{-3}$ \\
206317286.02\tablefootmark{(d)} &          & $0.711_{-0.025}^{+0.090}$ & $113.3327_{-0.0024}^{+0.0024}$ & $1.58252_{-0.00018}^{+0.00017}$ & $0.0133_{-0.0012}^{+0.0014}$ & $0.96_{-0.15}^{+0.11}$ & $0.57_{-0.24}^{+0.15}$ & 42 & 10.2 & 25.2 & 1.0 & $1.1~\times~10^{-2}$ & $4.6~\times~10^{-4}$ \\
212297394.02\tablefootmark{(g)} &          & $0.829_{-0.121}^{+0.087}$ & $363.4607_{-0.0020}^{+0.0020}$ & $2.28943_{-0.00019}^{+0.00019}$ & $0.0162_{-0.0010}^{+0.0010}$ & $1.323_{-0.095}^{+0.081}$ & $0.45_{-0.21}^{+0.16}$ & 33 & 11.1 & 24.2 & 1.0 & $4.4~\times~10^{-3}$ & $1.8~\times~10^{-4}$ \\
212424622.02\tablefootmark{(g)} &          & $1.106_{-0.120}^{+0.145}$ & $365.2616_{-0.0069}^{+0.0090}$ & $18.0983_{-0.0058}^{+0.0060}$ & $0.0148_{-0.0011}^{+0.0013}$ & $2.17_{-0.42}^{+0.67}$ & $0.74_{-0.07}^{+0.08}$ & 4 & 8.5 & 19.0 & 1.0 & $2.2~\times~10^{-2}$ & $9.2~\times~10^{-4}$ \\
212499991.02\tablefootmark{(g)} &          & $0.912_{-0.162}^{+0.037}$ & $382.6254_{-0.0096}^{+0.0081}$ & $34.885_{-0.010}^{+0.011}$ & $0.0193_{-0.0019}^{+0.0016}$ & $1.60_{-0.16}^{+0.13}$  & $0.28_{-0.19}^{+0.26}$ & 2 & 9.1 & 11.9 & 11.1 & $2.0~\times~10^{-1}$  & $9.9~\times~10^{-3}$ \\
212587672.02\tablefootmark{(fgk)} &          & $0.950_{-0.040}^{+0.040}$ & $375.4906_{-0.0054}^{+0.0053}$ & $15.2841_{-0.0029}^{+0.0037}$ & $0.0108_{-0.0007}^{+0.0007}$ & $1.12_{-0.09}^{+0.12}$ & $0.24_{-0.17}^{+0.25}$ & 5 & 8.6 & 18.5 & 23.5 & $1.3~\times~10^{-1}$ & $6.1~\times~10^{-3}$ \\
220554210.02\tablefootmark{(dk)} & K2-282    & $0.940_{-0.040}^{+0.040}$ & $532.1411_{-0.0017}^{+0.0017}$ & $0.70531_{-0.00005}^{+0.00005}$ & $0.0146_{-0.0007}^{+0.0007}$ & $1.48_{-0.10}^{+0.13}$ & $0.25_{-0.17}^{+0.24}$ & 103 & 16.8 & 25.8 & 1.0 & $4.0~\times~10^{-5}$ & $1.6~\times~10^{-6}$ \\
\hline
\end{tabular}
\tablefoot{$T_0$, $P$, $R_{\rm p}/R_{\rm s}$, and $b$ obtained from MCMC fitting. \#tr, S/N, and SDE$_{\rm TLS}$ obtained from {\tt TLS}. FPP$_2$ obtained with {\tt vespa} and corrected for the multiplicity boost following Eq.~\eqref{eq:fpp2}. $R_{\rm p}\,[R_\oplus]$ derived from $R_{\rm p}/R_{\rm s}$ as per {\tt TLS} and literature values for the stellar radius. All stellar radii and masses from \citet{2016ApJS..224....2H}, except for EPIC\,201497682\tablefootmark{(l)}, EPIC\,201615463\tablefootmark{(j)}, EPIC\,201754305\tablefootmark{(d)}, EPIC\,201841433\tablefootmark{(l)}, EPIC\,205950854\tablefootmark{(j)}, EPIC\,212297394\tablefootmark{(l)}, EPIC\,212499991\tablefootmark{(l)}, EPIC\,212587672\tablefootmark{(k)}, and EPIC\,220554210\tablefootmark{(k)}. \\
           \tablefoottext{a}{$T_0~=~{\rm BKJD}-2065$\,d with the Barycentric Kepler Julian Day BKJD = BJD - 2,454,833.0\,d.}\\
           \tablefoottext{b}{Number of transits detected with {\tt TLS} that have data.}\\
           Planets or candidates have previously been found in these systems by \tablefoottext{c}{\citet{2015ApJ...809...25M},} \tablefoottext{d}{\citet{2016ApJS..222...14V},} \tablefoottext{e}{\citet{2016ApJS..226....7C},} \tablefoottext{f}{\citet{2016A&A...594A.100B},} \tablefoottext{g}{\citet{2016MNRAS.461.3399P},} \tablefoottext{h}{\citet{2017AJ....154..207D},} \tablefoottext{i}{\citet{2018AJ....155..136M},} \tablefoottext{j}{\citet{2018AJ....156..277L},} \tablefoottext{k}{\citet{2018AJ....155...21P}.}\\
           Gaia DR2 radii were computed by \tablefoottext{l}{\citet{2018A&A...616A...8A}.}
            }
\end{sidewaystable*}

Our initial search in the 489 light curves revealed 50 new candidates that fulfilled our automated search criteria. For each candidate, we then inspected the respective vetting sheet by eye and thereby reduced the number of candidates to 20. We then inspected the {\it K2} aperture masks used by {\tt EVEREST} and compared them to high-resolution images from Pan-STARRS to evaluate the apparent separation of any possible contaminants. The resulting characterization was then submitted to {\tt vespa} for the calculation of the respective FPPs. This chain of successive vetting criteria resulted in the detection and statistical validation of 17 hitherto undetected planets.

Figure~\ref{fig:EPIC201754305} shows an example of our iterative transit search with {\tt TLS} that led us to the discovery of K2-16\,d (EPIC\,201754305.03). The top panel presents the {\it K2} light curve extracted with {\tt EVEREST} and the running median indicated with a red line, the center panel shows the detrended light curve used for the transit search with {\tt TLS}, and the three panels at the bottom illustrate the phase-folded light curves obtained from the respective $T_0$ and $P$ values detected with {\tt TLS} for each of the three planets. The new planet is shown in the bottom left panel.

The final sample of 17 validated planets was then characterized with the {\tt emcee} MCMC sampler. The results are listed in Table~\ref{tab:MCMC} including $T_0$, $P$, $R_{\rm p}/R_{\rm s}$, and $b$. Table~\ref{tab:MCMC} also summarizes the characteristics of the {\it K2} light curves as evaluated with {\tt TLS}, such as the number of detected transits for the new object (\#tr), S/N, and SDE$_{\rm TLS}$ as well as the quantities $\rho$, FPP, and FPP$_2$ that relate to the {\tt vespa} software.

In Fig.~\ref{fig:gallery} we present a gallery of the phase-folded transits of all 17 newly discovered transiting planets. Each panel contains the detrended and phase-folded {\it K2} light curve (black dots) and our best fit obtained through the MCMC sampling of the data (red line). The scales of the ordinates are different in the panels, which show transits depths ranging from about 100\,ppm (EPIC\,201615463.02) to more than 2000\,ppm (EPIC\,201238110.02). The very shallow depths of these phase-folded light curves, with S/Ns just between about 7 and 17 (see Table~\ref{tab:MCMC}), make it very hard for the human eye to classify the morphologies of the transit light curves as either box-like, transit-like, or V-shaped, etc. The question then arises why {\tt TLS} finds these noisy transits but previous studies that used box-fitting algorithms missed them. We have previously investigated this question and found three answers. First, {\tt TLS} is intrinsically more sensitive to shallow transits simply because it uses a more physical description of a transit light curve. This facts tends to produce higher SDE values for real planets and lower SDE values for false positives \citep{2019A&A...623A..39H}. Second, and different from the vast majority of {\tt BLS} implementations, the grid of trial periods used in {\tt TLS} is not linear. Instead, the grid of trial orbital frequencies is linear, which makes {\tt TLS} equally sensitive to short- and long-period transiting planets while minimizing the computational load \citep{2014A&A...561A.138O}. This suppresses the occurrence of aliases and false positives (\PaperI). Third, in \PaperI \, we showed that the Earth-sized planet K2-32\,e can indeed be found with {\tt BLS} in the {\it K2} data that were detrended from systematic effects by the {\tt EVEREST} pipeline, whereas previous studies used the slightly more noisy {\tt K2SFF} light curves \citep{2014PASP..126..948V}.


Figure~\ref{fig:architectures} is an illustration of the orbital architectures of the 17 systems with new planets. In each panel, we show the respective host star at the left and indicate its effective temperature by a color (see caption). Each panel shows all planets that are now known around these 17 stars. The positions of the planets are based on their transit impact parameters (along the ordinate) and orbital semi-major axes ($a$, along the abscissa) derived from our MCMC modeling of the entire light curve. For our estimations of $a$, we assumed Kepler's third law for circular orbits and neglected the planetary masses, $a=(GM_{\rm s}(P / 2\pi)^2)^{1/3}$, where $G$ is the gravitational constant. In each panel, the abscissa ranges from 0 to 0.2\,AU and the ordinate extends to one solar radius. All symbols are to scale with respect to the sizes of all objects shown.

In 11 out of 17 cases, the new planet that we found is now the innermost known planet in its system. In 13 out of 17 cases, the new planet is the smallest of the known planets in its system. In the remaining four cases, in which the new planet is not the smallest planet, it is the outermost of the known planets and therefore shows the smallest number of transits and, hence, the lowest SDE$_{\rm TLS}$ in this system.

In the period-radius diagram shown in Fig.~\ref{fig:period_radius} we place our sample of 17 new planets from {\it K2} in the broader context of the known population of transiting exoplanets. Different symbols denote planets from different samples. Open circles denote transiting planets that are not from {\it K2}, and thus mostly from the {\it Kepler} primary mission, blue circles refer to planet known from {\it K2}, and red circles depict planets from {\it K2} that were found by the {\tt TLS} Survey. The first statistically validated planet from the {\tt TLS} Survey, K2-32\,e (\PaperI), is denoted with an orange circle. Figure~\ref{fig:period_radius} translates the enhanced sensitivity of {\tt TLS} compared to {\tt BLS} for shallow transits, as found in simulations by \citet{2019A&A...623A..39H}, into the exoplanet period-radius diagram.

All of the new planets, except for one object that orbits a star larger than the sun, are smaller than two Earth radii, and as a consequence, produce very shallow transits. As shown in \PaperI \, for the case of K2-32\,e, these transits can be very close to the detection limit for a heuristically chosen SDE in {\tt BLS} (SDE$_{\rm BLS}$), but they can exceed an equivalent detection limit chosen for {\tt TLS} to produce the same false-positive rates \citep{2019A&A...623A..39H}. Consequently, we verify that {\tt TLS} is particularly suited for the search of small planets. The most remarkable object in terms of its small size is EPIC\,201497682.03, which we found to have a radius of just $0.692_{-0.048}^{+0.059}\,R_\oplus$, making it the second smallest planet ever discovered with {\it K2}. Another highlight is the super-Earth EPIC 201238110.02, which orbits its M dwarf host star in the habitable zone \citep{2013ApJ...765..131K}. This means it could potentially have liquid surface water, although we expect that the strong tidal forces from the star have aligned the spin with the orbit and therefore eroded any seasons on the planet \citep{2011A&A...528A..27H}.

\section{Discussion}

\begin{figure}[t]
\centering
\includegraphics[width=1.01\linewidth]{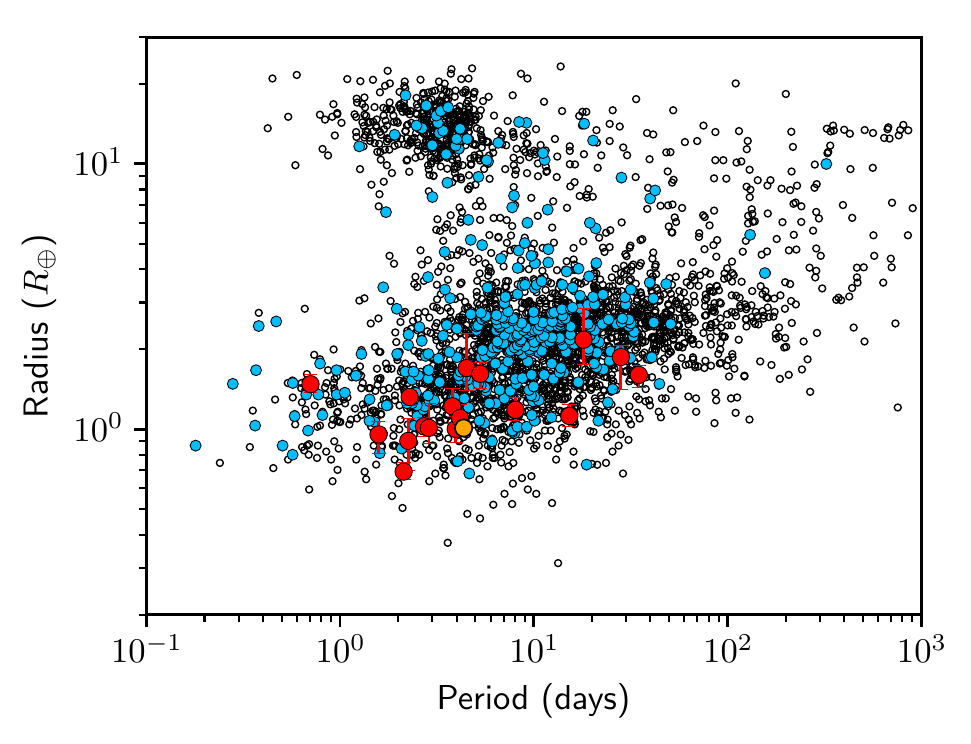}
\caption{Period-radius diagram of all confirmed transiting exoplanets (empty circles) mostly from K1, candidates and confirmed planets from {\it K2} (blue), and from the {\tt TLS} Survey (red), including K2-32\,e (orange) presented in \PaperI. The uncertainties in the orbital periods of the 18 planets discovered and characterized in the {\tt TLS} Survey are much smaller  and shorter than the symbol size.}
\label{fig:period_radius}
\end{figure}

The ability of {\tt TLS} to find hitherto unknown planets in the short-period Earth-sized regime (see Fig.~\ref{fig:period_radius}) is so high that a dedicated {\tt TLS}-based transit search has the potential of extending previous work on ultra-short-period planets  \citep{2013ApJ...779..165J,2014ApJ...787...47S,2016AJ....152...47A}. Short-period planets are particularly interesting from a planet-formation point of view, in particular in the context of the curious planet radius gap \citep{2017AJ....154..109F} and the stability of planetary atmospheres under extreme stellar irradiation \citep{2017ApJ...845..130L}. Small, ultra-short-period planets are also interesting in terms of their connection to stellar metallicity \citep{2017AJ....154...60W} and with regard to their tidal interaction with the star \citep{2015IJAsB..14..321B}.

A reanalysis of light curves that have previously been searched for with {\tt BLS} or similar algorithms is very likely to reveal new ultra-short-period Earth-sized planets. Our discovery rate of 18 newly validated planets (including K2-32\,e; \PaperI) around 489 stars with previously known planets ($18/489~{\approx}~3.7\,\%$) suggests the potential for the discovery of another roughly 100 additional planets in the data of the thousands of stars with planets and candidates from the {\it Kepler} primary mission. The numbers might be even higher given that the light curves from the {\it Kepler} primary mission contain four years of continuous observations compared to the approximately 80\,d covered by {\it K2}. Moreover, follow-up transit searches with {\tt TLS} can be used to validate known planet candidates by using the multiplicity boost in case of additional candidate detections, as done in this study.

\section{Conclusions}

We reported the discovery of 17 new planets in the archival data of the extended {\it Kepler} mission, {\it K2}. All of these planets are smaller than about $2.2\,R_\oplus$, and half of these planets are smaller than about $1.2\,R_\oplus$. While their extremely shallow transits were not securely detectable with standard transit-search algorithms looking for box-like transit signals, our newly developed {\tt TLS} transit search algorithm \citep{2019A&A...623A..39H} identified them as transiting candidates.

We used the {\tt vespa} software and optical images from Pan-STARRS to statistically validate these candidates with ${\rm FPP}~<~1\,\%$. Our characterization of these validated exoplanets based on an MCMC analysis of their entire {\it K2} light curves with the {\tt emcee} software reveals that one of our objects, EPIC\,201497682.03, with a radius of $0.692_{-0.048}^{+0.059}\,R_\oplus$ is now the second smallest planet ever discovered with {\it K2}.

Based on our discovery rate of one new planet around about 3.7\,\% of all stars from {\it K2} with previously know planets, we expect that {\tt TLS} can find another 100 small planets around the thousands of stars with planets and candidates from the {\it Kepler} primary mission that have been missed in previous searches.

\begin{acknowledgements}
The authors thank an anonymous referee for their comments that helped to clarify several passages in this manuscript. This research has made use of the Exoplanet Orbit Database and the Exoplanet Data Explorer at \href{http://exoplanets.org}{exoplanets.org}. This research has made use of the NASA Exoplanet Archive, which is operated by the California Institute of Technology, under contract with the National Aeronautics and Space Administration under the Exoplanet Exploration Program. This work made use of NASA's ADS Bibliographic Services. This research has made use of ``Aladin sky atlas'' developed at CDS, Strasbourg Observatory, France. The Pan-STARRS1 Surveys (PS1) and the PS1 public science archive have been made possible through contributions by the Institute for Astronomy, the University of Hawaii, the Pan-STARRS Project Office, the Max-Planck Society and its participating institutes, the Max Planck Institute for Astronomy, Heidelberg and the Max Planck Institute for Extraterrestrial Physics, Garching, The Johns Hopkins University, Durham University, the University of Edinburgh, the Queen's University Belfast, the Harvard-Smithsonian Center for Astrophysics, the Las Cumbres Observatory Global Telescope Network Incorporated, the National Central University of Taiwan, the Space Telescope Science Institute, the National Aeronautics and Space Administration under Grant No. NNX08AR22G issued through the Planetary Science Division of the NASA Science Mission Directorate, the National Science Foundation Grant No. AST-1238877, the University of Maryland, Eotvos Lorand University (ELTE), the Los Alamos National Laboratory, and the Gordon and Betty Moore Foundation. This work has made use of data from the European Space Agency (ESA) mission
{\it Gaia} (\href{https://www.cosmos.esa.int/gaia}{www.cosmos.esa.int/gaia}), processed by the {\it Gaia} Data Processing and Analysis Consortium (DPAC, \href{https://www.cosmos.esa.int/web/gaia/dpac/consortium}{www.cosmos.esa.int/web/gaia/dpac/consortium}). Funding for the DPAC has been provided by national institutions, in particular the institutions participating in the {\it Gaia} Multilateral Agreement. RH is supported by the German space agency (Deutsches Zentrum f\"ur Luft- und Raumfahrt) under PLATO Data Center grant 50OO1501. KR is a member of the International Max Planck Research School for Solar System Science at the University of G\"ottingen. KR performed the MCMC analysis of the light curves.
\end{acknowledgements}

\bibliographystyle{aa}
\bibliography{aa}

\end{document}